\documentclass[a4paper,12pt]{article}
\usepackage{amsfonts}
\usepackage{latexsym}
\usepackage{amssymb}
\usepackage{color}

           \newcommand{\be}{\begin{equation}}
            \newcommand{\ee}{\end{equation}}
            \newcommand{\bee}[1]{\begin{equation}\label{#1}}
            \newcommand{\bey}{\begin{eqnarray}}
            \newcommand{\byy}[1]{\begin{eqnarray}\label{#1}}
            \newcommand{\eey}{\end{eqnarray}}
            \newcommand{\R}[1]{(\ref{#1})}
            \newcommand{\C}[1]{\cite{#1}}

            \newcommand{\mvec}[1]{\mbox{\boldmath{$#1$}}}

            \newcommand{\q}{\st{\td}{\bf Q}}
            
            \newcommand{\w}{\st{\td}{\bf W}}
            \newcommand{\s}{\st{\td}{\bf S}}

            \newcommand{\E}{\underline{\bf 1}}

            \newcommand{\ro}{\stackrel{\ \circ}{\varrho}}

            \newcommand{\st}[2]{\stackrel{_#1}{#2}}
            \newcommand{\td}{{^{\bullet}}}
            \newcommand{\dm}{\diamond}
            \newcommand{\D}[1]{\st{\dm}{#1}}
            \newcommand{\trido}{\triangledown}

\begin{document}
\date{\empty}
\title{Phenomenological 
Quantum 
Thermodynamics\\of Closed Bipartite Schottky Systems
\thanks{In memory of Bob Axelrad}}
\author{W. Muschik\footnote{Corresponding author:
muschik@physik.tu-berlin.de}
\\
Institut f\"ur Theoretische Physik\\
Technische Universit\"at Berlin\\
Hardenbergstr. 36\\D-10623 BERLIN,  Germany}
\maketitle

\begin{abstract}\noindent
How to introduce thermodynamics to quantum mechanics ? Among from numerous
possibilities of solving this task, the simple choice is here: The conventional von 
Neumann equation deals with a density operator whose probability weights are
time independent. Because there is no reason apart from the reversible quantum
mechanics that these weights have to be time independent, this constraint is waived, thus
making possible to introduce thermodynamical concepts to quantum mechanics.
This procedure is similar to that of Lindblad's equation, but different on principle.
But beyond this simple starting-point, the applied thermodynamical concepts of discrete
systems may perform a "source theory" for other versions of phenomenological quantum
thermodynamics. 
\end{abstract}


\section{Introduction}

Conventional quantum mechanics is a reversible theory because the 
entropy production vanishes for all its processes. There are two 
possibilities to introduce irreversibility to quantum mechanics without 
restricting the full set of ob\-ser\-vables: One can change Schr\"{o}dinger's 
equation by introducing e.g. friction terms, an adventurous procedure which
we will not follow. Here we follow an other way which modifies von Neumann's
equation by introducing time dependent weights of the statistical operator, thus
generating irreversibility. That means, the in time changing
composition of the ensemble
which belongs to the statistical operator allows to introduce thermodynamical
concepts of irreversibility to quantum mechanics. 
This procedure is similar to that of Lindblad's equation, but different on principle,
because the entropy producing operators of the modified von Neumann equation
are generated by constitutive equations. Consequently, the dynamics of the considered
system is decomposed into a quantum mechanical part and a material dependent part
which is described beyond the Hamiltonians by constitutive equations.
Other possibilities
of implementing irreversibility,
introducing a restricted set of observables thus creating irreversibility by loss of
information \C{MUKA94}, or using statistical concepts such as microcanonical
or canonical ensembles as ad-hoc concepts
are out of scope of this paper. 

We consider Schottky systems, that are discrete systems\footnote{Discrete systems are
"boxes" which interact with their environment through "partitions" \C{MUASP,MU18}.} whose interaction
with their environment is macroscopically described by heat-, power- and material
exchange which here is especially suppressed by considering closed systems. Such a
discrete system can be described as an undecomposed or as a decomposed one.
Undecomposed means that the special structure of the interior of the Schottky system
is out of scope whereas the internal structure of the system is taken into account by a
decomposed description \C{MUBE04,MUBE07}. Here in quantum thermodynamics,
undecomposed means that the Hamiltonian of the system is not decomposed into the
partial Hamiltonians of the sub-systems and their interaction Hamiltonian. This case was
sketched in a previous paper \C{MU??}, whereas here the decomposed Schottky system
is treated, that means, the Hamiltonian of the system is decomposed into its parts.

The task of quantum thermodynamics is to determine quantum-theo\-re\-ti\-cally the
entropy, its time rate and production and the entropy exchange between system and
its environment and between its sub-systems. For this purpose, we start with a modified
von Neumann equation which allows to introduce a non-vanishing entropy rate by
time dependent weights of the statistical operator which are set to zero in conventional
quantum mechanics resulting in a non-thermal theory.

The paper is organized as follows: First of all, we remember the quantum
thermodynamics of undecomposed Schottky systems \C{MU??} in sect.\ref{US}
by using a modified von Neumann equation. Because non-equilibrium thermodynamics
works with a non-equilibrium temperature \C{MU77,MUBR77}, a short repetition
concerning contact temperature is given in sect.\ref{CT}. Sect.\ref{CS} deals in
detail with compound systems which are characterized by a decomposed Hamiltonian,
but by an undecomposed density operator. Equilibria, constitutive equations, adiabatic
and reversible processes are considered. The results achieved for compound systems
are transferred to decomposed systems in sect.\ref{DS}. Decomposed means, the
Hamiltonian as well as the density operator are decomposed into parts belonging to
sub-systems. A summary and two appendices finish the paper.

\section{Undecomposed Systems\label{US}}
\subsection{Schottky systems}

A discrete system $\cal G\subset\mathbb{R}$${^3}$, described as undecomposed and
homogeneous which is separated by a partition $\partial {\cal G}$ from its environment
${\cal G}^\Box$ is called a {\em Schottky system} \C{SCHO29}, if the interaction
between $\cal G$ and ${\cal G}^\Box$ through $\partial {\cal G}$ can be macroscopically
described by
\bee{+1}
\mbox{heat exchange}\ \st{\td}{Q},\quad\mbox{power exchange}\ \st{\td}{W},\quad
\mbox{and material exchange}\ \st{\td}{\mvec{n}}\!{^e}.
\ee
The power exchange is related to the work variables $\mvec{a}$ of the system
\bee{+2}
\st{\td}{W}\ =\ {\bf K}\cdot\st{\td}{\mvec{a}}.
\ee
Here, ${\bf K}$ are the generalized forces which are as well known as the work variables.
Kinetic and potential energy are constant and therefore out of scope. The heat exchange 
$\st{\td}{Q}$ is
measurable by calorimetry and the time rate of the mole numbers $\st{\td}{\mvec{n}}\!\!{^e}$
due to material exchange between $\cal G$ and ${\cal G}^\Box$ by weigh.

Here, we restrict ourselves to closed discrete systems for which the material exchange
is suppressed by a material impervious partition between the system and its environment
and chemical reactions are absent. The
Hamiltonian of an undecomposed closed system contains no interaction parts, and the
interaction between system and environment is macroscopically (semi-classically\footnote{Semi-classical means: heat- and power-exchange are not influenced by the
system's Hamiltonian.}) described by $\st{\td}{Q}$ and $\st{\td}{W}$.

\subsection{The modified von Neumann equation}

The undecomposed closed system is described by a density
operator
\bee{+3}
\varrho\ :=\ \sum_j p_j |\Phi^j \rangle\langle\Phi^j |,\quad 0\leq p_j \leq 1,
\quad \sum_j p_j = 1
\ee
which is composed of pure quantum states $\{|\Phi^j \rangle\}$ 
which are normalized, complete and orthogonal \C{KATZ67}
\bee{+4}
\langle\Phi^j |\Phi^j \rangle\ =\ 1,\quad \wedge k\neq l:\
\langle\Phi^k |\Phi^l \rangle= 0,\quad\sum_j |\Phi^j \rangle\langle\Phi^j |\ =\ \underline{1}.
\ee
Beyond that, the dynamics of these quantum states is given by the Schr\"odinger
equation
\bee{+5}
i\hbar\partial_t|\Phi^j \rangle\ =\ {\cal H}|\Phi^j \rangle,\qquad
j=1, 2, 3,... ,
\ee
in which the Hamiltonian is self-adjoint ${\cal H}^+={\cal H}$. Inserting \R{+5}
in \R{+3} results in the modified von Neumann equation \C{MU??,KATZ67}
\bee{+6}
\partial_t \varrho\ \equiv\ \st{\td}{\varrho}\ 
=\ -\frac{i}{\hbar}\Big[{\cal H},\varrho\Big]+\ro,\qquad
\ro\ :=\ \sum_j \st{\td}{p}_j|\Phi^j \rangle\langle\Phi^j |.
\ee
The modification is, that in contrast to the conventional quantum theory, the time rates of
the weights $\st{\td}{p}_j$ of the density operator \R{+3} do not all vanish

$\blacksquare$\underline{Setting I:}\footnote{
A "definition" is a formal short form for an expression without any physical background. A "setting" is a hypothesis induced by physics determining the axiomatic structure of a
theory, whereas "axiom" and "postulate" characterize established principles of formal
theories without any physical background.}
\bee{+7}
\vee\ j:\qquad\st{\td}{p}_j\ \neq\ 0\ \longrightarrow\ \ro\ \neq\ 0.
\hspace{4cm}\blacksquare\hspace{-2.6cm}
\ee

The modified von Neumann equation shows that the time dependence of the density
operator \R{+3} has two reasons: the quantum mechanical dyna\-mics represented by the 
commutator in \R{+6}$_1$ 
and the time dependence of the propagator $\ro$ in
\R{+6}$_2$. In conventional quantum theory of isolated systems, the
propagator does not appear,
because the $\{p_j\}$ are presupposed to be time independent, a fact 
which causes reversibility, as we will see below. 
Accepting setting I means, that irreversibility is generated
by an in time changing composition of the density operator $\varrho$.  

From \R{+3}, \R{+1} and \R{+6}$_2$ follows
\bee{+8}
\varrho |\Phi^k \rangle\ =\ p_k  |\Phi^k \rangle\quad
\mbox{Tr}\varrho\ =\ \sum_{j}{p}_j\ =\ 1,\quad
\mbox{Tr}\ro\ =\ \sum_{j} \st{\td}{p}_j\ =\ 0.
\ee

\subsection{The first law}

The Hamiltonian  ${\cal H}$ belongs to a non-isolated closed undecomposed system:
no material exchange and missing chemical reactions, but power- and heat-exchange
between the system and its  environment. The energy of the considered system is 
\bee{+9}   
E\ :=\ \mbox{Tr}({\cal H}\varrho),
\ee
and the time rate of the energy
\bee{+10}
\st{\td}{E}\ =\ \mbox{Tr}(\st{\td}{\cal H}\varrho) + 
\mbox{Tr}({\cal H}\st{\td}{\varrho})
\ee
can be split into power- and heat-exchange according to the 1st 
law of thermodynamics of closed systems
\bee{+11} 
\st{\td}{W}\ :=\ \mbox{Tr}(\st{\td}{\cal H}\varrho),\qquad
\st{\td}{Q}\ :=\ \mbox{Tr}({\cal H}\st{\td}{\varrho}).
\ee
Inserting the modified von Neumann equation \R{+8} into \R{+11}$_2$, we obtain
after a short, but simple calculation \C{MU??} 
\bee{+12}
\st{\td}{Q}\ =\ \mbox{Tr}\Big({\cal H}\ro\Big)\ =\ 
\mbox{Tr}\Big({\cal H}\sum_j\st{\td}{p}_j|\Phi^j \rangle\langle\Phi^j |\Big),
\ee
and by introducing suitable work variables $\mathbf{a}$, we obtain
the power exchange \R{+11}$_1$
\bee{+13}
\st{\td}{W}\ =\ \mbox{Tr}\Big(\frac{\partial{\cal H}}{\partial\mathbf{a}}
\varrho\Big)\cdot\st{\td}{\mvec{a}}\ =:\ \mathbf{K}\cdot\st{\td}{\mvec{a}} .
\ee
Here, $\mathbf{K}$ are the so-called generalized forces.

\subsection{Entropy time rate}

The non-equilibrium entropy $S$ of the considered system, the Shannon 
entropy, is introduced \C{1KATZ67} as a
\vspace{.3cm}\newline
$\blacksquare$\underline{Setting II:}
\bee{+14}
S(\varrho)\ :=\ -k_B \mbox{Tr} (\varrho\ln\varrho)\hspace{3.5cm}\blacksquare
\hspace{-4cm}
\ee
($k_B\ =\ $ Boltzmann constant).
\vspace{.3cm}\newline
Inserting the modified von Neumann equation \R{+6}$_1$ into the Shannon 
non-equilibrium entropy \R{+14}, we obtain after a short calculation \C{MU??} by
taking \R{+6} and \R{+8}$_1$ into account
\bey\nonumber
\st{\td}{S}(\varrho,\st{\td}{\varrho}) &=& 
-k_B \mbox{Tr}(\ro\ln\varrho)\ =\ 
-k_B \mbox{Tr}\Big(\ro\ln(Z\varrho)\Big)\ =\ 
\\ \label{+15}
&=&-\sum_j\st{\td}{p}_j\langle\Phi^j|k_B \ln(Z\varrho)\Phi^j \rangle\ =:\
-\st{\td}{\mathbf p}\cdot{\mathbf f}^I,
\\ \label{+16}
f_k^I &:=& \langle\Phi^k|k_B \ln(Z\varrho)\Phi^k\rangle\ =\  k_B\ln(Zp_k),
\quad \wedge Z\in R^1_+,\hspace{.3cm}
\eey
Using \R{+8}$_3$, the second equal sign in \R{+15} follows from
\bee{+17}
 \mbox{Tr}\Big(\ro\ln(Z\varrho)\Big)\ =\ 
\mbox{Tr}\Big(\ro([\ln Z]\E+\ln\varrho)\Big)\ =\ \mbox{Tr}(\ro\ln\varrho).
\ee
Consequently, the entropy rate $\st{\td}{S}$ does not depend on $Z$ according to
\R{+17}. Therefore, $Z$ can be chosen arbitrarily below in connection with equilibrium.
Beyond that, $\st{\td}{S}$ does not depend on the Hamiltonian.

\subsection{Entropy exchange and production}

In this easy case of a closed non-isolated undecomposed system,
the entropy exchange is defined by multiplying the heat exchange with the
reciprocal of the contact temperature $\Theta$ which is discussed at some length
in sect.\ref{CT}. Starting with \R{+12}$_2$, we obtain an expression which looks formally
like the entropy rate \R{+15}$_4$
\bey\nonumber
\Xi &:=& \frac{\st{\td}{Q}}{\Theta}\ =\
\mbox{Tr}\Big(\frac{{\cal H}}{\Theta}\sum_j\st{\td}{p}_j|\Phi^j\rangle
\langle\Phi^j |\Big)\ =
\\ \label{+18} 
&=& \sum_j\st{\td}{p}_j\langle\Phi^j |\frac{{\cal H}}{\Theta}\Phi^j \rangle\ =:\ 
\st{\td}{\mathbf p}\cdot{\mathbf f}^{II},\quad f^{II}_k\ :=\ 
\langle\Phi^k |\frac{{\cal H}}{\Theta}\Phi^k \rangle.\hspace{.5cm}
\eey

The entropy rate \R{+15} is thermodynamically decomposed into the entropy exchange 
$\Xi$ and the entropy production $\Sigma$ \C{MU93}
\bee{+19}
\st{\td}{S}\ =\ \Xi + \Sigma.
\ee
Consequently, the entropy production is according to \R{+15}$_4$ and \R{+19}
\bee{+20}
\Sigma\ =\ \st{\td}{S} - \Xi\ =\ -\st{\td}{\mathbf p}\cdot\Big({\mathbf f}^{I}+
{\mathbf f}^{II}\Big)\ =:\ -\st{\td}{\mathbf p}\cdot\mathbf f.
\ee
The conclusions which result from \R{+15}, \R{+18} and \R{+20} for undecomposed
systems are discussed in detail in \C{MU??}. Here we are interested in decomposed
systems which differ from undecomposed ones by a different shape of the
Hamiltonian containing
an interaction term according to the decomposition of the system into sub-systems. 
But before discussing compound and decomposed
systems in sect.\ref{CS} and  sect.\ref{DS}, some properties of the
non-equilibrium contact temperature are presented in the next section.

\section{Contact Temperature\label{CT} and Internal Energy}

The thermal description of compound sytems in non-equilibrium requires a
concept of temperature beyond the thermostatic equilibrium temperature
as the following example demonstrates: two systems of different thermostatic
temperatures are in thermal contact which each other. The resulting compound
system is in non-equilibrium, and a joint thermostatic temperature for it does not exist.
Looking for a coarse description of this compound system --we do not ask for the
origin of the non-equilibrium (here the heat conduction)-- we need a non-equilibrium
analogue of the thermostatic temperature, that is the contact temperature
which is discussed in the sequel.

The non-equilibrium contact temperature $\Theta$  in \R{+18} is defined for closed systems
without chemical reactions by the inequality \C{MU77,MUBR77,MUBE07,MU09,MU18a,MU14} 
\bee{+21}
\st{\td}{Q}\Big(\frac{1}{\Theta}-\frac{1}{T^\Box}\Big)\ \geq\ 0
\ee
as follows: the closed system is contacted with an equilibrium environment ${\cal G}^\Box$ 
of the thermostatic temperature $T^\Box$ generating the net heat exchange
$\st{\td}{Q}$ between the system and its environment. For defining the contact
tempe\-rature, we choose a special equilibrium environment of the thermostatic
temperature $T^\Box_\odot$ so that the net heat exchange vanishes
$\st{\td}{Q}=0$. If according to \R{+21}, the heat exchange has a change of sign
\bee{+22}
\varepsilon,\ \eta >0,\qquad T^\Box_\odot \pm\varepsilon\ \longrightarrow\  
\st{\td}{Q}\ =\ \pm\eta(\varepsilon),
\ee
then $T^\Box_\odot$ is by definition the contact temperature of the non-equilibrium
system
\bee{+23}
\Theta\ :=\ T^\Box_\odot\ \longrightarrow\ \st{\td}{Q}=0
\ee
and according to \R{+22}, the bracket in \R{+21} is continuous at $\st{\td}{Q}=0$.
Now using the
\vspace{.3cm}\newline
$\blacksquare$\underline{Proposition} \C{MU85}
\bey\nonumber
{\bf X}\cdot f({\bf X})\ \geq\ {\bf 0}\ (\mbox{for all}\ {\bf X}\wedge f \ \mbox{continuous at}\ 
{\bf X} = {\bf 0})\ \Longrightarrow\ 
\\ \nonumber
\Longrightarrow\ f({\bf 0}) = {\bf 0}\ \longleftrightarrow\  f({\bf X})\ =\ 
 {\bf M}({\bf X})\cdot{\bf X},
\\ \label{+24}
 {\bf M}({\bf X})\ \mbox{positive semi-definite}.
\hspace{.8cm}\blacksquare\hspace{-1cm}
\eey
By use of the setting\footnote{$\st{*}{=}$ marks a setting}
\bee{24a}
{\bf X}\ \st{*}{=}\ \Big(\frac{1}{\Theta}-\frac{1}{T^\Box}\Big),\qquad
 f({\bf X})\ \st{*}{=}\ \st{\td}{Q},
\ee
we obtain from \R{+24} the constitutive equation of the heat exchange including a
material dependent "heat conduction" $\kappa$
\bee{+25}
\st{\td}{Q}\ =\ \kappa\Big(\frac{1}{\Theta}-\frac{1}{T^\Box}\Big),\qquad
\kappa\Big[\frac{1}{\Theta}-\frac{1}{T^\Box}\Big]\ >\ 0.
\ee
In words, we created the following statement:
\begin{center}
\parbox[t]{11cm}{
{\sf Definition:} The system's contact temperature $\Theta$ is that thermostatic temperature
of the system's
equilibrium environment $T^\Box$ for which the net heat exchange $\st{\td}{Q}$ between the
system and this environment through an inert partition\footnotemark\ vanishes by change
of sign.} 
\end{center}\footnotetext{inert means, the partition
does not emit or absorb power, heat and material}
The non-equilibrium contact temperature is a state variable of the system and the question
arises: what is the connection between contact temperature and (internal) energy ? 

As easily demonstrated, contact temperature $\Theta$ and the
energy $E$
of a closed discrete system are
independent of each other. For this purpose, a rigid inert partition $\partial {\cal G}$
($\st{\td}{\mvec{a}} \equiv\mvec{0}$) and a time-dependent environment
temperature $T^\Box(t)$ is chosen which is always set equal to the value of the momentary
contact temperature $\Theta(t)$ of the closed system. We obtain
according to the first law \R{+10} and \R{+11} and an inert partition between system
and environment
\bee{+26}
T^\Box (t)\st{*}{=}\Theta (t)\ \longrightarrow\ 
\st{\td}{Q}\ =\ 0\ \longrightarrow\ \st{\td}{E}\ =\ 0.
\ee
Because $\Theta$ is time-dependent and $E$ is constant, totally
different from thermostatics, both quantities are independent of each other.

A quantum theoretical definition of the contact temperature is given in sect.\ref{2LAW}.

\section{Compound Systems\label{CS}}

First of all, we have to define the following expressions:\begin{center}
undecomposed, compound and decomposed description\end{center}
of a Schottky system. {\em Undecomposed systems} which were discussed
in sect.\ref{US} are not decomposed into sub-systems, whereas {\em compound} and
{\em decomposed systems} are composed of sub-systems. Compound systems are
described by only one joint density operator $\varrho_{com}$ which belongs to both
sub-systems, whereas in decomposed systems each sub-system
has its particular density operator, $\varrho^1$ and $\varrho^2$ .
Consequently, the description of undecomposed and compound
systems is similar because of their single density operator, whereas the description of decomposed
systems is more complex than that of undecomposed and compound systems. Considering
{\em bipartite systems} in the following sections, the density operators and the Hamiltonians
are as follows:
\bey\nonumber
\mbox{undecomposed:}&\quad&\varrho,\ {\cal H},\hspace{5.15cm}
\mbox{sect.\ref{US}} 
\\ \nonumber
\mbox{compound:}&\quad&\varrho_{com},\  {\cal H}=
{\cal H}^1+{\cal H}^2+{\cal H}^{12},
\hspace{1.35cm}\mbox{sect.\ref{CS}}
\\ \nonumber
\mbox{decomposed:}&\quad&\varrho^1,\ \varrho^2,\  {\cal H}=
{\cal H}^1+{\cal H}^2+{\cal H}^{12},\hspace{1.1cm} \mbox{sect.\ref{DS}}
\eey

\subsection{Partial Hamiltonians}

In bipartite compound systems, the Hamiltonian ${\cal H}$ of the undecomposed system
is decomposed into the sum of the partial Hamiltonians of the two 
sub-systems, ${\cal H}^1$ and ${\cal H}^2$, and of the interaction Hamiltonian
${\cal H}^{12}$ describing the interaction between these two
sub-systems\footnote{Using the semi-classical description, ${\cal H}^{12}$ does not contain any interaction between the bipartite system and its environment} 
\bee{+37}
{\cal H}\ =\ {\cal H}^1 + {\cal H}^2 + {\cal H}^{12}. 
\ee
With respect to the tensorial base \R{+27}, we used  in \R{+37} the abbreviations
\bee{+38}
{\cal H}^1 \otimes I^2\ \equiv\ {\cal H}^1,
\qquad
I^1\otimes{\cal H}^2\ \equiv\ {\cal H}^2,
\quad\longrightarrow\quad
[{\cal H}^1,{\cal H}^2]\ =\ 0,
\ee
with the unity operators $I^i,\ i=1,2,$ belonging to the corresponding factors of 
the tensor product \R{+27}. The interaction of one sub-system with the environment
is for the present described {\em semi-classically}, that means, by power- and
heat-exchanges which result from the partial Hamiltonians ${\cal H}^1$ and
${\cal H}^2$. The interaction Hamiltonian ${\cal H}^{12}$ refers only to the 
interaction between the two sub-systems and is independent of the system's
environment\footnote{We will get rid of the semi-classical description in sect.\ref{RSCD}}.

\subsection{Tensorial density operator}

The undecomposed system together with
its density operator $\varrho$ \R{+3}$_1$ and its propagator $\ro$ \R{+6}$_2$
is divided into two sub-systems by choosing 
a basis $\{|\Psi^k_1\rangle\}$ belonging to sub-system \#1 and an other one 
$\{|\Psi^l_2\rangle\}$ belonging to the other sub-system \#2.
The tensor product of these bases form an orthogonal basis of the compound system
\bee{+27}
\{|\Psi^k_1\rangle\otimes
|\Psi^l_2\rangle\}\ \equiv\ \{|\Psi^k_1\rangle|\Psi^l_2\rangle\}.
\ee
The pure quantum states $|\Phi^j\rangle$ in \R{+4} of the undecomposed system are
replaced by pure tensorial quantum states of the compound system
\bee{+28}
|\Phi^j\rangle\ \longrightarrow\ |\Phi^{kl}_{12}\rangle\ \equiv\ 
(|\Psi_1^k\rangle|\Psi_2^l\rangle),\quad\wedge k,l
\ee
Consequently, we define the density operator and the propagator of the
compound system according to \R{+3}$_1$ and \R{+6}$_2$
\bee{+29}
\varrho_{com}\ :=\ \sum_{kl} p_{kl} |\Phi^{kl}_{12} \rangle\langle\Phi^{kl}_{12}|,\quad
\ro_{com}\ :=\ \sum_{kl} \st{\td}{p}_{kl}|\Phi^{kl}_{12} \rangle\langle\Phi^{kl}_{12}|,
\ee
and like \R{+8} and \R{+16}
\bee{§29}
\varrho_{com}|\Phi^{rs}_{12} \rangle\ =\ p_{rs} |\Phi^{rs}_{12} \rangle.
\ee
The time derivative of the density operator \R{+29}$_1$ is
\bey\nonumber
\st{\td}{\varrho}_{com}\ =\ \ro_{com}+\sum_{kl}p_{kl}
\Big\{
(|\Psi_1^k\rangle^\td|\Psi_2^l\rangle)(\langle\Psi_2^l|\langle\Psi_1^k|)+
\\ \nonumber
+(|\Psi_1^k\rangle|\Psi_2^l\rangle^\td)(\langle\Psi_2^l|\langle\Psi_1^k|)+
\\ \nonumber
+(|\Psi_1^k\rangle|\Psi_2^l\rangle)(\langle\Psi_2^l|^\td\langle\Psi_1^k|)+
\\ \label{§29a}
+(|\Psi_1^k\rangle|\Psi_2^l\rangle)(\langle\Psi_2^l|\langle\Psi_1^k|^\td)
\Big\}
\eey
We now presuppose the validity of the Schr\"odinger equation for $|\Psi^m_X\rangle$ taking
into account the partial interaction operators --${\cal H}^{12}_1$ and
${\cal H}^{12}_2$-- between the two sub-systems
\bee{§5}
|\Psi_1^k\rangle^\td\ =\ -\frac{i}{\hbar}({\cal H}^1+{\cal H}^{12}_1)|\Psi_1^k\rangle,\qquad
|\Psi_2^l\rangle^\td\ =\ -\frac{i}{\hbar}({\cal H}^2+{\cal H}^{12}_2)|\Psi_2^l\rangle.
\ee
Consequently, \R{§29a} results in
\bee{+43za}
\st{\td}{\varrho}_{com}\ =\ \ro_{com} 
-\frac{i}{\hbar} \Big[({\cal H}^1+{\cal H}^{12}_1),\varrho_{com}\Big]
-\frac{i}{\hbar} \Big[({\cal H}^2+{\cal H}^{12}_2),\varrho_{com}\Big].
\ee
Introducing the total Hamiltonian
\bee{§43z}
{\cal H}\ :=\ {\cal H}^1+{\cal H}^{12}_1+{\cal H}^2+{\cal H}^{12}_2,
\ee
\R{+43za} results in
\bee{+43z}
\st{\td}{\varrho}_{com}\ =\ 
-\frac{i}{\hbar} \Big[{\cal H},\varrho_{com}\Big]+\ro_{com}.
\ee
The interaction Hamiltonian is composed of the partial interaction operators
\bee{43zb}
{\cal H}^{12}\ =\ {\cal H}^{12}_1+{\cal H}^{12}_2
\ee
which operate on the quantum states \R{+28} of the corresponding sub-systems.

\subsection{The exchanges}

From \R{+11}$_1$ and \R{+12} follows the power and heat exchange:\footnote{for
simplicity, we use the same symbols for undecomposed and
compound systems: $\st{\td}{W}_{com}\longrightarrow\st{\td}{W}$}
\bee{+30}
\st{\td}{W} =
\mbox{Tr}\Big(\st{\td}{{\cal H}}\sum_{kl}{p}_{kl}|\Phi^{kl}_{12} \rangle\langle\Phi^{kl}_{12}|\Big),\ \
\st{\td}{Q} =
\mbox{Tr}\Big({\cal H}\sum_{kl}\st{\td}{p}_{kl}|\Phi^{kl}_{12} \rangle\langle\Phi^{kl}_{12}|\Big),
\ee
and from \R{+15} and \R{+16} the entropy time rate:
\bee{+31}
\st{\td}{S}\ =\ 
-\sum_{kl}\st{\td}{p}_{kl}\langle\Phi^{kl}_{12}|k_B \ln(Z\varrho_{com})\Phi^{kl} _{12}\rangle\ =:\
-\st{\td}{\mathbf p}\cdot{\mathbf f}^I,
\ee
with the abbreviation
\bee{+32}
f_{kl}^I\ =\ \langle\Phi^{kl}_{12}|k_B \ln(Z\varrho_{com})\Phi^{kl} _{12}\rangle\ =\
k_B\ln(Zp_{kl}).
\ee
According to \R{+18}, the entropy exchange is
\byy{+33}
&&\Xi\ =\ \mbox{Tr}\Big(\frac{{\cal H}}{\Theta}\sum_{kl}
\st{\td}{p}_{kl}|\Phi^{kl}_{12}\rangle
\langle\Phi^{kl}_{12} |\Big)\ =\ \st{\td}{\mathbf p}\cdot{\mathbf f}^{II},
\\ \label{+34}
&&f^{II}_{kl}\ :=\ \langle\Phi^{kl}_{12} |\frac{{\cal H}}{\Theta}\Phi^{kl}_{12} \rangle.
\eey
The density operator \R{+29}$_1$, the propagator \R{+29}$_2$ and the entropy time
rate \R{+31} do not depend on the Hamiltonian or on the contact temperature, whereas
the power exchange \R{+30}$_1$ and 
the heat exchange \R{+30}$_2$ depend on the Hamiltonian and the entropy exchange
\R{+33} additionally on the contact temperature.
According to \R{+20}, the entropy production is
\byy{+35}
\Sigma\ =\ \st{\td}{S} - \Xi\ =\ -\st{\td}{\mathbf p}\cdot
\Big({\mathbf f}^I+{\mathbf f}^{II}\Big)\
=:\  -\st{\td}{\mathbf p}\cdot{\mathbf f},
\\ \label{+36}
 f_{kl}\ :=\ \langle\Phi^{kl}_{12} |\Big(k_B\ln(Z\varrho_{com})
+\frac{{\cal H}}{\Theta}\Big)\Phi^{kl}_{12} \rangle.
\eey

\subsection{Externally isolated and non-isolated systems\label{IS}}

Evidently, the temporal progress is different in isolated and non-isolated systems.
The act of isolating the system 
is achieved by introducing an insulating partition between the
system and its environment which does not influence the state of the system,
but all its time rates.
These time rates in consideration are
$\st{\td}{S}$, $\Xi$, $\Sigma$ and $\st{\td}{\mathbf p}$, and they change by
isolating the system ($\longrightarrow$) into 
\byy{36aQ}
\st{\td}{S} &\longrightarrow& \st{\td}{S}_{iso},\qquad\st{\td}{S}_{ex}\ :=\
\st{\td}{S} -\st{\td}{S}_{iso},
\\ \label{36bQ}
\Xi &\longrightarrow& \Xi_{iso},\qquad\Xi_{ex}\ :=\ \Xi - \Xi_{iso},
\\ \label{36cQ}
\Sigma &\longrightarrow& \Sigma_{iso},\qquad \Sigma_{ex}\ :=\ 
\Sigma - \Sigma_{iso},
\\ \label{36dQ}
\st{\td}{\mathbf p} &\longrightarrow& \st{\td}{\mathbf p}\!{^{iso}},\qquad\st{\td}{\mathbf p}\!{^{ex}}\ :=\ \st{\td}{\mathbf p} - \st{\td}{\mathbf p}\!{^{iso}}.
\eey
The so called external time rates $\boxplus_{ex}$ which represent the jumps induced by isolating the system are defined by \R{36aQ}$_2$ to \R{36dQ}$_2$.

Consequently, the time rates decompose considering \R{+31}, \R{+33} and \R{+35}
\byy{36a}
\st{\td}{S} &=& \st{\td}{S}_{ex}+\st{\td}{S}_{iso}\ =\
-\st{\td}{\mathbf p}\cdot{\mathbf f}^I,
\\ \label{36b}
\Xi &=& \Xi_{ex}+\Xi_{iso}\ =\ \st{\td}{\mathbf p}\cdot{\mathbf f}{^{II}},
\\ \label{36c}
\Sigma &=& \Sigma_{ex} + \Sigma_{iso}\ =\ - \st{\td}{\mathbf p}\cdot{\mathbf f},
\\ \label{36d}
\st{\td}{\mathbf p} &=& \st{\td}{\mathbf p}\!{^{ex}}+\st{\td}{\mathbf p}\!{^{iso}}.
\eey

In more detail, we accept the following
\vspace{.3cm}\newline
$\blacksquare$\underline{Setting III \C{MU??}:}\newline
The installation of an isolating partition between the bipartite system ${\cal G}$
and its environment ${\cal G}^\Box$ does not change the partial
Hamiltonians\footnote{${\cal H}^{12}$ describes the interaction between the
sub-systems. That between system and environment is for the present described
semi-classically by
heat- and power-exchange without any Hamiltonian.} and the density operator of the
system in semi-classical description.
This isolation influences the propagator, that means, the time rate of the density
operator transforms by changing the time rates of its weights
\bee{+47}
\st{\td}{\mathbf p}\quad \longrightarrow\quad  \st{\td}{\mathbf p}\!{^{iso}}.
\hspace{4.5cm}\blacksquare\hspace{-4.5cm}
\ee

Evident is that the exchange quantites $\st{\td}{W}$ and $\st{\td}{Q}$ depend
also on the state of the system's environment. According to \R{+13}, the power
exchange is controlled by the rates of the work variables $\st{\td}{\mvec{a}}$,
whereas the heat exchange \R{+25}$_1$ is controlled by the difference of the contact
temperature and the thermostatic temperature of the controlling equilibrium environment
according to \R{+25}. Consequently, the rates of the weights $\st{\td}{\mathbf p}$ of the density operator depend on the $\st{\td}{W}$ and $\st{\td}{Q}$, controlling
quantities in semi-classical description.

When isolating the bipartite system from its 
environment, these rates change according to setting III \R{+47}\footnote{This "jumping"
of the propagator by an external isolation of the system is similar to the "reduction of the wave function" in quantum mechanics}. According to \R{36d}
\bee{+48}
\st{\td}{\mathbf{p}}\!{^{ex}}\ :=\ \st{\td}{\mathbf{p}}\ -\ \st{\td}{\mathbf{p}}\!{^{iso}},
\quad\longrightarrow\quad\st{\td}{\mathbf{p}}\!{^{ex,iso}}\ =\ \mathbf{0},
\ee
we obtain for the time rates of the weights of the density operator in
\byy{+49}
\mbox{non-isolated closed systems:}&\qquad&
\st{\td}{\mathbf{p}}\ =\ \st{\td}{\mathbf{p}}\!{^{iso}}
+\st{\td}{\mathbf{p}}\!{^{ex}},
\\ \label{+50}
\mbox{isolated systems:}&\qquad&\st{\td}{\mathbf{p}}\!{^{iso}}.
\eey
The quantities ${\mathbf f}^I$ and ${\mathbf f}^{II}$, \R{+32} and \R{+34},
are independent of an external isolation of the system.

We obtain from \R{+30}$_2$ and \R{+49} the split into an {\em exchange propagator} 
$\ro_{ex}$ and a dissipative thermal part  $\ro_{iso}$, called the 
{\em irreversibility propagator}, 
\byy{+51}
\ro_{com}\ = \ \ro_{ex} + \ro_{iso},\hspace{4.6cm}
\\ \label{+52}
\ro_{ex}\ :=\ \sum_{kl} \Big\{\st{\td}{p}_{kl}\!{^{ex}}|
\Phi^{kl}_{12}\rangle\langle\Phi^{kl}_{12}|\Big\},\quad
\ro_{iso}\ :=\ \sum_{kl} \Big\{\st{\td}{p}_{kl}\!{^{iso}}
|\Phi^{kl}_{12}\rangle\langle\Phi^{kl}_{12}|\Big\}.
\eey

The splitting into external and internal quantities in \R{36a} to \R{36d}
allows to formulate the following
\vspace{.3cm}\newline
$\blacksquare$\underline{Setting IV:}\newline 
The entropy production is not influenced by isolating the system:
\bee{IV1}
\Sigma\ \equiv\ \Sigma_{iso}\ \longrightarrow\ \Sigma_{ex}\ \equiv\ 0.
\ee
An isolation of the system induces that the entropy exchange vanishes:
\bee{IV2}
\Xi_{iso}\ \equiv\ 0\ \longrightarrow\ \Xi\ \equiv\ \Xi_{ex}.
\hspace{3.4cm}\blacksquare\hspace{-3.4cm}
\ee

Taking \R{+35}$_3$ and \R{+49} into account, \R{IV1}$_1$ results in
\bee{IV3}
-\st{\td}{\mathbf{p}}\cdot{\mathbf f}\ =\
-\st{\td}{\mathbf{p}}{^{iso}}\cdot{\mathbf f}\ =\ 
-\st{\td}{\mathbf{p}}{^{ex}}\cdot{\mathbf f}-\st{\td}{\mathbf{p}}{^{iso}}\cdot{\mathbf f}\ \longrightarrow\
\st{\td}{\mathbf{p}}{^{ex}}\cdot{\mathbf f}\ \equiv\ 0.
\ee
Taking \R{+33}$_2$ and \R{+49} into account, \R{IV2}$_2$ results in
\bee{IV4}
\st{\td}{\mathbf{p}}\cdot{\mathbf f}{^{II}}\ =\ 
\st{\td}{\mathbf{p}}{^{ex}}\cdot{\mathbf f}{^{II}}\ =\
\st{\td}{\mathbf{p}}{^{ex}}\cdot{\mathbf f}{^{II}}+
\st{\td}{\mathbf{p}}{^{iso}}\cdot{\mathbf f}{^{II}}\ \longrightarrow\
\st{\td}{\mathbf{p}}{^{iso}}\cdot{\mathbf f}{^{II}}\ \equiv\ 0.
\ee
Using \R{IV3}$_3$ and \R{IV4}$_3$, we obtain from \R{+35} and \R{+33}
\bee{58c}
\Sigma\ =\ -\st{\td}{\mathbf p}\cdot{\mathbf f}\ =\
-\st{\td}{\mathbf p}\!{^{iso}}\cdot{\mathbf f}{^I},\qquad
\Xi\ =\ \st{\td}{\mathbf p}\!{^{ex}}\cdot{\mathbf f}^{II}\ =\
-\st{\td}{\mathbf p}\!{^{ex}}\cdot{\mathbf f}^{I}.
\ee

\subsubsection{Power exchange}

The work variables in \R{+13} belong to the undecomposed system. Switching over to a
compound system, these work variables have to be replaced by those which belong
to the sub-systems --$\mvec{a}^1, \mvec{a}^2$-- and  those which are
related to the interaction between them $\mvec{a}^{12}$
\bee{38d}
\mvec{a}\ \longrightarrow\ (\mvec{a}^1, \mvec{a}^2, \mvec{a}^{12})
\longrightarrow\ 
{\cal H}(\mvec{a}^1, \mvec{a}^2, \mvec{a}^{12}).
\ee
According to their definition, the work variables of the compound system \R{38d}$_2$
are attached to the partial Hamiltonians as follows\footnote{
$\mvec{a}^{12}$
describes the position of a partition between the sub-systems which is displacable thus
influencing the tree Haniltonians.}
\bee{38d1}
{\cal H}^1(\mvec{a}^1,\mvec{a}^{12}),\quad {\cal H}^2(\mvec{a}^2,\mvec{a}^{12}),\quad {\cal H}^{12}(\mvec{a}^{12}),
\ee
resulting in
\byy{38d2}
\st{\td}{\cal H}\!{^1}\ =\ \frac{\partial{\cal H}^1}{\partial\mvec{a}^1}
\cdot\st{\td}{\mvec{a}}\!{^1}
+\frac{\partial{\cal H}^1}{\partial\mvec{a}^{12}}
\cdot\st{\td}{\mvec{a}}\!{^{12}},
&&
\st{\td}{\cal H}\!{^2}\ =\ \frac{\partial{\cal H}^2}{\partial\mvec{a}^2}\cdot\st{\td}{\mvec{a}}\!{^2}
+\frac{\partial{\cal H}^2}{\partial\mvec{a}^{12}}
\cdot\st{\td}{\mvec{a}}\!{^{12}},
\\ \label{38d3}
\st{\td}{\cal H}\!{^{12}} &=& \frac{\partial{\cal H}^{12}}{\partial\mvec{a}^{12}}
\cdot\st{\td}{\mvec{a}}\!{^{12}}.
\eey
According to \R{+30}$_1$,
\bee{38a1a}
\mbox{Tr}\Big(\st{\td}{\cal H}\!{^A}\varrho_{com}\Big)\ =:\ \st{\td}{W}\!{^A},
\qquad A=1,2,12,
\ee
this decomposition of the time derivative of the Hamiltonian
allows to define external and internal power exchanges: external between the
sub-systems and the environment of the compond system and internal exchanges
between the sub-systems themselves
\byy{38d4}
\st{\td}{W}\!{{^A}\!\!_{ex}} &:=&
\mbox{Tr}\Big(\frac{{\partial\cal H}^A}{\partial\mvec{a}^A}\varrho_{com}\Big)
\cdot\st{\td}{\mvec{a}}\!{^A},\quad A=1,2,\quad
\frac{{\partial\cal H}^{12}}{\partial\mvec{a}^A}\ =\ \mvec{0},
\\ \label{38d5}
\st{\td}{W}\!{{^A}\!\!_{int}} &:=&
\mbox{Tr}\Big(\frac{{\partial\cal H}^A}{\partial\mvec{a}^{12}}\varrho_{com}\Big)
\cdot\st{\td}{\mvec{a}}\!{^{12}},\quad A=1,2,12.
\eey
Consequently, we obtain
\bee{38d6}
\st{\td}{W}_{ex}\ =\ \st{\td}{W}\!{{^1}\!\!_{ex}}+\st{\td}{W}\!{{^2}\!\!_{ex}},
\qquad
\st{\td}{W}_{int}\ =\ \st{\td}{W}\!{{^1}\!\!_{int}}+\st{\td}{W}\!{{^2}\!\!_{int}}+
\st{\td}{W}\!{{^{12}_{int}}}.
\ee

Accepting that the sum of the internal power exchanges is zero, we have the following 
\vspace{.3cm}\newline
$\blacksquare$\underline{Setting V:}
\bee{38e}
\st{\td}{W}_{int}\ =\ \mbox{Tr}\Big(\frac{\partial\cal H}{\partial\mvec{a}^{12}}\varrho_{com}\Big)
\cdot\st{\td}{\mvec{a}}\!{^{12}}\equiv\ 0,
\hspace{2.8cm}
\blacksquare\hspace{-2.8cm}
\ee
and we obtain from \R{38d6}$_2$
\bee{38e1}
-\st{\td}{W}\!{^1_{int}}\ =\ \st{\td}{W}\!{^2_{int}}+
\st{\td}{W}\!{^{12}_{int}}.
\ee
If $\st{\td}{W}\!{{^{12}}\!\!_{int}}\rangle0$, the partition between the sub-systems
is power absorbing and if $\st{\td}{W}\!{{^{12}}\!\!_{int}}\langle0$ power supplying.

Formally, five cases appear, if $\st{\td}{W}\!{^{12}_{int}}$ is zero
\byy{38e2}
\st{\td}{W}\!{{^{12}_{int}}}\ = &0& =\ 
\mbox{Tr}\Big(\frac{{\partial\cal H}^{12}}{\partial\mvec{a}^{12}}\varrho_{com}\Big)
\cdot\ \st{\td}{\mvec{a}}\!{^{12}}\ \longrightarrow\ 
\\ \label{38e2a}
&\longrightarrow&\mbox{1)}\quad\st{\td}{\mvec{a}}\!{^{12}}=\mvec{0}\quad
\wedge\quad\st{\td}{W}\!{{^{A}_{int}}}\ =\ 0,
\\ \label{38e2b}
&\longrightarrow&\mbox{2)}\quad
{\cal H}^{12}\ =\underline{0}\quad\wedge\quad
\st{\td}{W}\!{{^{A}_{int}}}\ \neq\ 0,
\\ \label{38e2c}
&\longrightarrow&\mbox{3)}\quad
\frac{{\partial\cal H}^{12}}{\partial\mvec{a}^{12}}\ =\ \mvec{0}\ \longrightarrow\ 
\st{\td}{\cal H}\!{^{12}}\ =\ \underline{0},
\\ \label{38e2d}
&\longrightarrow&\mbox{4)}\quad
\mbox{Tr}\Big(\frac{{\partial\cal H}^{12}}{\partial\mvec{a}^{12}}\varrho_{com}\Big)
\ =\ \mvec{0},
\\ \label{38e2e}
&\longrightarrow&\mbox{5)}\quad
\mbox{Tr}\Big(\frac{{\partial\cal H}^{12}}{\partial\mvec{a}^{12}}\varrho_{com}\Big)
\perp\ \st{\td}{\mvec{a}}\!{^{12}}.
\eey
Here, we only consider the cases $\#1$ and $\#2$ which demonstrate that an internal
power exchange can appear,  even if the interaction Hamiltonian ${\cal H}^{12}$
vanishes because the ${\cal H}^A$ depend on $\st{\td}{\mvec{a}}\!{^{12}}$
according to \R{38d2}.

\subsubsection{Heat exchange}

Similar to the Hamiltonians, there are also three contact temperatures: $\Theta^1$ and
 $\Theta^2$ belonging to the sub-systems, and $\Theta^{12}$ --for the
present an
unknown contact temperature-- which belongs to the quantum mechanical interaction.
Besides these three contact temperatures of the bipartite system, according to \R{+21},
an additional contact temperature $\Theta$ of the undecomposed description of the
compound system exists.

According to \R{+11}$_2$, we define the partial heat- and entropy-exchanges
by taking the equation of motion \R{+6} and the decomposition of the Hamiltonian
\R{+37} into account
\bey\nonumber
\st{\td}{Q}\!{^A} &:=& \mbox{Tr}\Big({\cal H}^A\st{\td}{\varrho}_{com}\Big)\ =
\\ \label{38a}
&:=&-\frac{i}{\hbar}\mbox{Tr}\Big({\cal H}^A\Big[{\cal H}^{12},\varrho_{com}\Big]\Big)+
\mbox{Tr}\Big({\cal H}^A\ro_{com}\Big),\quad\mbox{A=1,2,}
\\ \nonumber
\st{\td}{Q}\!{^{12}} &:=& \mbox{Tr}\Big({\cal H}^{12}\st{\td}{\varrho}_{com}\Big)\ =
\\ \label{38a1}
&:=&-\frac{i}{\hbar}\mbox{Tr}\Big({\cal H}^{12}\Big[({\cal H}^{1}+{\cal H}^{2}),\varrho_{com}\Big]\Big)+\mbox{Tr}\Big({\cal H}^{12}\ro_{com}\Big).
\\ \label{38a2}
\Xi^A &:=& \frac{\st{\td}{Q}\!{^A}}{\Theta^A}\quad\quad\mbox{A=1,2,12}.
\eey
The partial heat- and work-exchanges are additive according to \R{+37}, whereas the
entropy exchanges are not additive because of the different contact temperatures of the
sub-systems and the partition between them: 
\bee{38b}
\st{\td}{Q}\!{^1} + \st{\td}{Q}\!{^2} +\st{\td}{Q}\!{^{12}}\ =\ \st{\td}{Q},\qquad
\frac{\st{\td}{Q}\!^1}{\Theta^1}+\frac{\st{\td}{Q}\!^2}{\Theta^2}+\frac{\st{\td}{Q}\!^{12}}{\Theta^{12}}\ \geq\ \frac{\st{\td}{Q}}{\Theta},
\ee
A proof of the inequality \R{38b}$_2$ can be found in \C{MUBE04} and more
detailled in app.\ref{EEI} \R{a11}. The fact, that the entropy exchange of the
undecomposed system is not equal
to the sum of the entropy exchanges of the compound system according to \R{38b}$_2$, is called the {\em compound deficiency} of the entropy exchanges \C{MUBE04,MUBE07}.
Compound deficiencies will be treated in more detail in sect.\ref{CD}.

The power exchange was decomposed into its external and its internal part according to
\R{38d6} and \R{38e}. A similar decomposition is true for the heat- and
entropy-exchange which is based on the following statement:
the sum of the internal heat exchanges vanishes and each partial internal heat exchange
vanishes with vanishing quantum mechanical interaction. We now consider the sum
of the first terms of \R{38a}$_3$ and \R{38a1}$_3$
\bey\nonumber
-\frac{i}{\hbar}\sum_{A=1,2}\mbox{Tr}\Big({\cal H}^A\Big[{\cal H}^{12},\varrho_{com}\Big]\Big)
-\frac{i}{\hbar}\mbox{Tr}\Big({\cal H}^{12}\Big[({\cal H}^{1}+{\cal H}^{2}),\varrho_{com}\Big]\Big)\ =\ 0\ =
\\ \label{38b1}
= -\frac{i}{\hbar}\sum_{A=1,2}^{12}\mbox{Tr}\Big({\cal H}^A\Big[{\cal H},\varrho_{com}\Big]\Big)
\eey
which vanishes according to \R{+37}. Beyond that, each term of \R{38b1} vanishes
with vanishing ${\cal H}^{12}$. Consequently, we accept according to \R{38a} and
\R{38a1} the following
\vspace{.3cm}\newline
$\blacksquare$\underline{Setting VI:}
\bee{38b2}
\st{\td}{Q}{^A_{int}}\ :=\ 
-\frac{i}{\hbar}\mbox{Tr}\Big({\cal H}^A\Big[{\cal H},\varrho_{com}\Big]\Big)
+\mbox{Tr}\Big({\cal H}^A\ro{_{iso}}\Big),\quad\mbox{A=1,2,12}.
\hspace{.7cm}\blacksquare\hspace{-.7cm}
\ee

Because
\bee{38b3}
\st{\td}{Q}{^A}\ =\ \st{\td}{Q}{^A_{ex}}+\st{\td}{Q}{^A_{int}},
\quad\mbox{A=1,2,12},
\ee
the external heat exchanges are according to \R{38a} and \R{38a1}
\bee{38b4}
\st{\td}{Q}{^A_{ex}}\ =\ \mbox{Tr}\Big({\cal H}^A\ro{_{ex}}\Big),
\quad\mbox{A=1,2,12}.
\ee
the corresponding entropy exchanges are according to \R{38a1} and \R{38b2}
\byy{38b5}
\Xi^A_{ex}\ &=& -\frac{i}{\hbar}\mbox{Tr}\Big(
\frac{{\cal H}^A}{\Theta^A}\ro{_{ex}}\Big)\ =\
\st{\td}{\mathbf p}\cdot{\mathbf f}^{IIA},
\\ \label{38b6}
f^{IIA} &:=& \langle\Phi^k_{12} |\frac{{\cal H}^A}{\Theta^A}\Phi^k_{12} \rangle
\quad\mbox{A=1,2,12}.
\\ \label{38b7}
\Xi^A_{int} &=& -\frac{i}{\hbar}\mbox{Tr}\Big(
\frac{{\cal H}^A}{\Theta^A}\Big[{\cal H},\varrho_{com}\Big]\Big)
+\mbox{Tr}\Big(\frac{{\cal H}^A}{\Theta^A}\ro{_{iso}}\Big),
\quad\mbox{A=1,2,12}.
\eey
According to \R{38b1}, we obtain from \R{38b2}
\bee{38b8}
\sum_{A=1,2}^{12}\st{\td}{Q}{^A_{int}}\ =\ 0\ \longrightarrow\ 
\mbox{Tr}({\cal H}\ro_{iso})\ \equiv\ 0,\ \wedge\
\mbox{Tr}\Big(\frac{{\cal H}}{\Theta}\ro_{iso}\Big)\ \equiv\ 0.
\ee

According to \R{38b}, \R{38b1} and \R{38b2}, the heat- and entropy-exchanges satisfy 
\byy{+59}
\st{\td}{Q}_{ex} &=& \st{\td}{Q}\!{^1_{ex}} + \st{\td}{Q}\!{^2_{ex}}
+\st{\td}{Q}{_{ex}^{12}},
\\ \label{+60}
0 &=& \st{\td}{Q}\!{^1_{int}} + \st{\td}{Q}\!{^2_{int}}
+\st{\td}{Q}\!{^{12}_{int}},\quad\st{\td}{Q}_{int}\ \equiv\ 0,
\\ \label{+61}
{\Xi}_{ex} &\leq& {\Xi}{^1_{ex}} + {\Xi}{^2_{ex}}
+\Xi^{12}_{ex},
\\ \label{+62}
0 &\leq& {\Xi}{^1_{int}} + {\Xi}{^2_{int}} +{\Xi}{^{12}_{int}},\hspace{.6cm}
\Xi_{int}\ \equiv\ 0.
\eey
The decomposition into the external and internal parts is achieved for the power
exchanges by the work variables according to \R{38d4} and \R{38d5}. With respect
to the heat- and entropy-exchanges, this decomposition is not determined by the work
variables but by the thermodynamical properties of the internal heat exchanges \R{38b2}.

\subsubsection{Entropy rate and 2$^{\bf nd}$ law\label{2LAW}}

As already mentioned, the entropy rate \R{+15}$_2$ of the compound system depends
neither on the Hamiltonian nor on the contact temperatures and is therefore not influenced
by the decomposition of the Hamiltonian
\bee{+41}
\st{\td}{S}\ =\
-k_B \mbox{Tr}\Big(\st{\td}{\varrho}_{com}\ln(Z\varrho_{com})\Big)\ =\
-k_B \mbox{Tr}\Big(\ro_{com}\ln(Z\varrho_{com})\Big).
\ee
According to \R{+19}, \R{IV1}, \R{IV2} and \R{36a}, the entropy rate results in
\bee{+74}
\st{\td}{S}\ =\ \Sigma + \Xi\ =\ \Sigma_{iso} + \Xi_{ex}\ =\
\Sigma_{iso} + \frac{\st{\td}{Q}_{ex}}{\Theta}ü\ =\ 
\st{\td}{S}_{iso}+\st{\td}{S}_{ex}.
\ee
Consequently, we obtain a quantum mechanical expression of the contact temperature
of the compound system taking \R{+41} and \R{38b4} into account
\bee{+77}
\frac{1}{\Theta}\ =\ \frac{\st{\td}{S}_{ex}}{\st{\td}{Q}_{ex}}\ =\
\frac{-k_B\mbox{Tr}\Big(\ro_{ex}\ln(Z\varrho_{com})\Big)}
{\mbox{Tr}({\cal H}\ro_{ex})}.
\ee

The sum of the entropy exchanges is according to \R{38b4}, \R{38b} and \R{+61}
\bee{+42}
\sum_{A=1,2}^{12}\Xi^A_{ex}\ =\
-\frac{i}{\hbar}\mbox{Tr}\Big(\sum_{A=1,2}^{12}\frac{{\cal H}^A}{\Theta^A}
\ro_{ex}\Big)\ \leq\ 
-\frac{i}{\hbar}\mbox{Tr}\Big(\frac{{\cal H}}{\Theta}\ro_{ex}\Big).
\ee

Introducing the\vspace{.3cm}\newline
$\blacksquare$\underline{Second Law:}
\bee{+75}
\Sigma\ \st{*}{\geq}\ 0,\hspace{5cm}\blacksquare\hspace{-5cm}
\ee
and taking \R{+21}\ into account, \R{+74}$_2$ results in\footnote{considering cyclic
processes, result in Clausius' inequality and its extension including the contact
temperature}
\bee{+76}
\st{\td}{S}\ \geq\ \st{\td}{S}_{ex}\ =\ \frac{\st{\td}{Q}_{ex}}{\Theta}\ 
\geq\ \frac{\st{\td}{Q}_{ex}}{T^\Box}.
\ee

A similar representation as \R{+76}$_2$ of the contact temperatures of the sub-systems,
$\Theta^1$ and $\Theta^2$, is not possible because partial entropies of the sub-systems
are not defined for compound systems with respect to the joint propagator and
density operator \R{+29}. The entropy rate \R{+31} does not depend on the Hamiltonian
or its decomposition. Partial entropies can be defined for decomposed systems which
are discussed in sect.\ref{DS}

The entropy production \R{58c}$_1$ is not negative according to the Second Law
\R{+75}
\bee{*76} 
\Sigma\ =\ -\st{\td}{\mathbf p}\cdot{\mathbf f}\ =\
-\st{\td}{\mathbf p}\!{^{iso}}\cdot{\mathbf f}^I\ \geq\ 0.
\ee
But pay attention to the fact that different expressions of the entropy production belong to
different controlled systems: \R{*76}$_1$ belongs according to \R{+49} to a non-isolated system, whereas \R{*76}$_2$ represents the entropy production of an isolated
system according to \R{+50}. The value of the entropy production is equal in both
systems because isolation of a system does not influence the entropy production according
to \R{IV1}$_1$. This difference of the meaning of \R{*76}$_1$ and \R{*76}$_2$ comes
into sight, if equilibria are discussed sect.\ref{EQUI} . 

The propagator $\ro_{com}$ vanishes in conventional quantum mechanics and with
it also the entropy rate and entropy exchange and production according to \R{+41}
and \R{+42}. The entropy itself is constant in time according to \R{+15}, and the heat
exchange \R{38a} vanishes in conventional quantum mechanics. That is the reason why
conventional quantum mechanics is characterized as an adiabatical and reversible theory.

As in \C{MU??}, we consider special thermodynamical
processes and their quantum theoretical interpretations. The well-known 
phenomenological concepts of adiabatic, 
irreversible and reversible processes in isolated and in closed
systems and the concept of equilibrium are interpreted 
quantum-theoretically without using methods of statistical thermodynamics: 
we are looking for a {\em phenomeno\-lo\-gical irreversible quantum 
thermodynamics}.

\subsection{Vanishing quantum mechanical interaction}
\subsubsection{Internal exchanges}

If the quantum mechanical interaction vanishes, we obtain from \R{38b2} by taking
\R{+60} and \R{+62} into accout\footnote{concerning power exchange, see \R{38e2}}, 
\byy{+63}
\mbox{if}\ 
{\cal H}^{12}=0 &\longrightarrow& \st{\td}{Q}\!{^{12}_{int}} = 0 \longrightarrow\ 
\st{\td}{Q}\!{^1_{int} = -\st{\td}{Q}\!{^2_{int}}},
\\ \label{+64}
&\longrightarrow& \Xi{^{12}_{int}} = 0 \longrightarrow\ 
\Xi{^1_{int} \geq -\Xi{^2_{int}}}.
\eey
According to \R{+63}$_3$, the partition between the two sub-systems is
inert\footnote{ inert means: the partition does not emit or absorb neither heat nor power}
and heat- and power-exchange through this partition are continuous, if the 
quantum mechanical interaction vanishes. Or in other words, the quantum mechanical
interaction causes according to \R{38e1} and \R{+60} that the partition between the
sub-systems is not inert, that means, it absorbs or emits $\st{\td}{Q}\!{^{12}_{int}}$ or/and $\st{\td}{W}\!{^{12}_{int}}$, representing a "moving heat absorbing partition
with friction".

Presupposing that the entropy exchange through the partition between the sub-systems is
continuous, we obtain by use of \R{+60} 
\bee{64a} 
0\ \st{*}{=}\ 
\frac{\st{\td}{Q}\!{^1_{int}}}{\Theta^{1}}+\frac{\st{\td}{Q}\!{^2_{int}}}{\Theta^{2}}\ =\ 
\frac{\st{\td}{Q}\!{^1_{int}}}{\Theta^{1}}-\frac{\st{\td}{Q}\!{^1_{int}}+
\st{\td}{Q}\!{^{12}_{int}}}{\Theta^{2}}\ =\ \st{\td}{Q}\!{^1_{int}}\Big(
\frac{1}{\Theta^{1}}-\frac{1}{\Theta^{2}}\Big)-
\frac{\st{\td}{Q}\!{^{12}_{int}}}{\Theta^{2}},
\ee
resulting in the {\em condition of continuous entropy exchange}
\bee{64b}
\frac{\st{\td}{Q}\!{^{12}_{int}}}{\Theta^{2}}\ =\ 
\st{\td}{Q}\!{^1_{int}}\Big(\frac{1}{\Theta^{1}}-\frac{1}{\Theta^{2}}\Big).
\ee
If the quantum mechanical interaction vanishes, and consequently the heat exchange is continuous according to \R{+63}$_3$,
\bee{+65}
{\cal H}^{12}\ =\ 0:\hspace{1.5cm}
0\ =\ \st{\td}{Q}\!{^1_{int}}\Big(\frac{1}{\Theta^{1}}-\frac{1}{\Theta^{2}}\Big)
\ee
follows by taking \R{+63}$_2$ into account.
Because the contact temperatures of the two sub-systems are presupposed to be different,
we obtain from \R{+65}, \R{+63} and \R{+64}
\bee{+66}
{\cal H}^{12}=0\ \wedge\ \Theta^1\neq\Theta^2\ \longrightarrow\ \st{\td}{Q}\!{^1_{int}}=0\
=\ \st{\td}{Q}\!{^2_{int}}\ =\ \Xi{^2_{int}}\ =\ \Xi{^1_{int}}.
\ee 
that means, we have proved the following\vspace{.3cm}\newline
$\blacksquare$\underline{Proposition:}
\begin{enumerate} 
\item If the quantum mechanical interaction does not vanish ${\cal H}^{12}\ \neq\ 0$:
\vspace{-.3cm}
\begin{itemize}
\item the heat exchange through the partition between the sub-systems is not continuous,
\item the corresponding entropy exchange is continuous, if the continuity condition \R{64b} is satisfied.
\end{itemize}
\item If the quantum mechanical interaction vanishes ${\cal H}^{12}\ =\ 0$:\vspace{-.3cm}
\begin{itemize}
\item  the heat exchange through the partition between the sub-systems is continuous:
the partition is inert \R{+63}$_2$,
\item the corresponding entropy exchange is not continuous \R{+64}$_2$,
\item heat exchange and entropy exchange are jointly continuous only in the trite case that the partition between the two sub-systems is insulating and consequently the
internal exchanges vanish \R{+66}.\hfill$\blacksquare$
\end{itemize}\end{enumerate}

\subsubsection{External exchanges}

In this section, we consider the following problem: A bipartite compound system is in contact
with a macroscopic equilibrium environment of the thermostatic temperature $T^\Box$.
The interaction of the bipartite system with the environment is semi-classically described by 
heat- and entropy-exchanges $\st{\td}{Q}\!\!{_{ex}^B},\  \Xi{_{ex}^B},\ B=1,2$, according to \R{+59} and \R{+61}.
According to \R{+21}, the contact temperatures $\Theta^B$ of the sub-systems satisfy the
following inequalities
\bee{+69}
\st{\td}{Q}\!\!{_{ex}^B}\Big(\frac{1}{\Theta^B}-\frac{1}{T^\Box}\Big)\ \geq\ 0,\quad B=1,2.
\ee

Presupposing
\bee{69a}
{\cal H}^{12}\ =\ 0\quad\longrightarrow\quad\st{\td}{Q}\!\!{_{ex}^{12}}\ =\ 0,
\quad\Xi{_{ex}^{12}}\ =\ 0,
\ee
and using the contact temperature $\Theta$ of the undecomposed system, we obtain according to \R{+61} and \R{+59}
\bee{+70}
{\Xi}{^1_{ex}} + {\Xi}{^2_{ex}}\ =\ 
\frac{\st{\td}{Q}\!\!{_{ex}^1}}{\Theta^1}+\frac{\st{\td}{Q}\!\!{_{ex}^2}}{\Theta^2}\ \geq\ 
\Xi_{ex}\ =\ \frac{\st{\td}{Q}\!\!{_{ex}}}{\Theta},\qquad
\st{\td}{Q}\!\!{_{ex}^1}+\st{\td}{Q}\!\!{_{ex}^2}\ =\ \st{\td}{Q}\!\!{_{ex}},
\ee
resulting in
\byy{+71}
\st{\td}{Q}\!\!{_{ex}^1}\Big(\frac{1}{\Theta^1}-\frac{1}{T^\Box}\Big)+
\st{\td}{Q}\!\!{_{ex}^2}\Big(\frac{1}{\Theta^2}-\frac{1}{T^\Box}\Big) &\geq&
\st{\td}{Q}\!\!{_{ex}}\Big(\frac{1}{\Theta}-\frac{1}{T^\Box}\Big)\ \geq\ 0,
\\ \label{71a}
\longrightarrow\hspace{.4cm}
\frac{\st{\td}{Q}\!\!{_{ex}^1}}{\Theta^1}+\frac{\st{\td}{Q}\!\!{_{ex}^2}}{\Theta^2}
&\geq& \frac{\st{\td}{Q}\!\!{_{ex}}}{T^\Box}, 
\eey
an inequality which demonstrates the compatibility of the contact temperatures
$\Theta^1$ and $\Theta^2$ of the sub-systems with $\Theta$, the contact temperature
of the undecomposed system. The relations presented in app.\ref{EEI} are
satisfied by taking the entropy exchange inequality \R{38b}$_2$ into account.

\subsection{Equilibria\label{EQUI}}

\subsubsection{Equilibrium conditions}

Equilibria are defined by the following equilibrium conditions
\byy{*77}
\st{\td}{\varrho}{^{eq}_{com}}\ \st{*}{=}\ 0,\quad\wedge\quad
\ro{^{eq}_{ex}}\ \st{*}{=}\ 0\quad\wedge\quad\ro{^{eq}_{iso}}\ \st{*}{=}\ 0,
\\ \label{*77a}
\st{\td}{\mvec{a}}\!{^A_{eq}}\ \st{*}{=}\ {\mathbf{0}}, \quad \mbox{A=1,2,12},\qquad
\Theta^A_{eq}\ \st{*}{=}\ \Theta_{eq}.
\eey
From \R{+43z} and \R{*77} follows
\bee{*78}
 [{\cal H},\varrho{^{eq}_{com}}]\ =\ 0, 
\ee
and from \R{38d4}, \R{38d5} and \R{*77a}$_1$
\bee{*78a}
\st{\td}{W}{^{Aeq}_{ex}}\ =0, \quad\mbox{A=1,2},\qquad
\st{\td}{W}{^{Aeq}_{int}}\ =0, \quad\mbox{A=1,2,12}.
\ee
According to \R{*77}$_{2,3}$ and \R{*78}, the heat- and entropy-exchanges \R{38b4}, \R{38b5} and \R{38b2}, \R{38b7} vanish in equilibrium
\byy{+79}
&&\st{\td}{Q}\!{^{Aeq}_{ex}}\ =\ 0,\qquad\Xi^{Aeq}_{ex}\ =\ 0,\quad\mbox{A=1,2,12},
\\ \label{79a}
&&\st{\td}{Q}\!{^{Aeq}_{int}}\ =\ 0,\qquad\Xi^{Aeq}_{int}\ =\ 0,\quad\mbox{A=1,2,12}.
\eey
An other shape of \R{*77}$_{2,3}$ is
\bee{+81}
\st{\td}{\mathbf{p}}\!{_{eq}^{ex}}\ 
=\ {\mathbf{0}}\quad \wedge\quad \st{\td}{\mathbf{p}}\!{^{iso}_{eq}}\ =\ \mathbf{0}.
\ee
Consequently, the entropy rate \R{+41}$_1$ and the entropy production \R{*76} vanish in equilibrium
\bee{81a}
\st{\td}{S}\!{^{eq}}\ =\ 0,\qquad\Sigma^{eq}\ =\ 0.
\ee

\subsubsection{Equilibrium distributions}

We now consider the following situation: a system (isolated or non-isolated) is in equilibrium. According to \R{*76}, we obtain for the two different kinds of systems
\bee{81b}
\Sigma_{eq}\ =\ 0\ \longrightarrow\ 0\ =\
\st{\td}{\mathbf p}\!{^{iso}_{eq}}\cdot{\mathbf f}^{I}_{eq},
\quad\vee\quad 0\ =\ \st{\td}{\mathbf p}_{eq}\cdot{\mathbf f}_{eq}.
\ee
An external manipulation brings the system out of equilibrium by
changing the equilibrium condition \R{+81} of the propagator arbitrarily\footnote{a more detailled argumentation can be found in sect.\ref{CE}}
\bee{81c}
\st{\td}{\mathbf{p}}\!{_{eq}^{ex}}\ \longrightarrow\
\st{\td}{\mathbf{p}}\!{_{\#}^{ex}}
\quad \vee\quad 
\st{\td}{\mathbf{p}}\!{_{eq}^{iso}}\ \longrightarrow\ 
\st{\td}{\mathbf{p}}\!{^{iso}_{\#}}
\ee
which does not influence ${\mathbf f}^{Ieq}$ and ${\mathbf f}^{eq}$
according to \R{+32} and \R{+36}. Because of \R{*76}, the replacement
\R{81c} generates
\bee{81d}
\Big(0\ \leq\
-\st{\td}{\mathbf p}\!{^{iso}_{\#}}\cdot{\mathbf f}^{I}_{eq}
\quad\vee\quad 0\ \leq\ -\st{\td}{\mathbf p}_{\#}\cdot{\mathbf f}_{eq}\Big),
\quad\mbox{for all }
\Big(\st{\td}{\mathbf{p}}\!{^{iso}_{\#}},\st{\td}{\mathbf{p}}\!{_{\#}^{ex}}\Big).
\ee
Because $\st{\td}{\mathbf p}\!{^{iso}_{\#}}$ and $\st{\td}{\mathbf p}_{\#}$ can be
chosen arbitrarily\footnote{more details in sect.\ref{CE}}, the only possibility to satisfy \R{81d} is
\byy{+82}
{\mathbf f}{^{I}_{eq}}\ =\ \mathbf{0},\quad&& \mbox{in isolated systems} 
\\ \label{+83}
{\mathbf f}{_{eq}}\ =\ \mathbf{0},\quad&& \mbox{in non-isolated closed systems}.
\eey
Consequently, \R{+82} and \R{+83} result in\footnote{see \R{+32}}
\byy{+84}
f_{kl}^{Ieq} &=& \langle\Phi^{kl}_{eq} |k_B \ln(Z\varrho^{eq}_{com})\ \Phi^{kl}_{eq} \rangle\ 
=\ 0\ =\ k_B\ln(Zp_{kl}^{eq}), \quad\wedge kl,
\\ \label{+85}
f_{kl}^{eq} &=& \langle\Phi^{kl}_{eq} |\Big(k_B \ln(Z\varrho^{eq}_{com}) 
+ \frac{{\cal H}}{\Theta_{eq}}\Big)\Phi^{kl}_{eq} \rangle\ =\ 0,\quad\wedge kl.
\eey
The equilibrium density operator $\varrho^{eq}_{com}$ follows from \R{+84} for
isolated systems and from \R{+85} for non-isolated closed systems, as derived in
the next two sections.
Catchword like, one can say: The Second Law applied to equilibrium generates the
equilibrium distributions.

\subsubsection{Isolated systems: micro-canonical ensemble}

From \R{+84}$_3$ follows
\bee{+86}
Zp_{kl}^{eq}\ =\ 1\ \longrightarrow\ p_{kl}^{eq}\ =\ \frac{1}{Z}\ 
\longrightarrow\ \varrho^{eq}_{com}\ =\ \frac{1}{Z}\sum_{kl}
|\Phi^{kl}_{eq}\rangle\langle\Phi^{kl}_{eq}|\ =\ \frac{1}{Z}{\underline 1}.
\ee
Tracing the density operator according to \R{+8}$_2$ results in
\bee{+87}
1\ =\ \sum_{kl} \langle\Phi^{kl}_{eq}|\varrho^{eq}_{com}\Phi^{kl}_{eq}\rangle\ 
=\ \frac{1}{Z}\sum_{kl}\langle\Phi^{kl}_{eq}|\Phi^{kl}_{eq}\rangle\ =\
\frac{1}{Z}\sum_{kl} 1_{kl}.
\ee
The last sum must be restricted because of convergency: $1\leq kl\equiv J\leq N$. 
Consequently, according to \R{+87}$_3$
$Z=N$ is valid, and the density operator \R{+86}$_3$ of an 
isolated system has, as expected, the micro-canonical form
\bee{5.1bx}
\varrho_{mic}\ =\ \frac{1}{N}\sum_{J=1}^N |\Phi^J_{eq}\rangle\langle{\Phi}^J_{eq}|,\qquad
N<\infty.
\ee

\subsubsection{Non-isolated closed systems: canonical ensemble}

According to \R{*78}, the Hamilton operator commutes with the density ope\-rator
in equilibrium. Consequently, a common system of eigenfunctions exists for both
operators, and we presuppose that this system is given by \R{§29}. Consequently, from \R{+85} follows by taking \R{+84} into account
\bey\nonumber
 k_B\ln(Zp_{kl}^{eq}) + \frac{E_{kl}}{\Theta_{eq}}\ =\ 0\ \longrightarrow\hspace{6.5cm}
\\ \label{I3}
\longrightarrow\ k_B
\sum_{kl}\ln(Zp_{kl}^{eq})|\Phi^{kl}_{eq}\rangle\langle{\Phi}^{kl}_{eq}| 
+\sum_{kl} |\Phi^{kl}_{eq}\rangle\frac{E_{kl}}{\Theta_{eq}}\langle{\Phi}^{kl}_{eq}|\ =\ \underline{0},
\eey 
resulting according to \R{+32} in
\bee{I4}
k_B\ln(Z\varrho^{eq}_{com}) + \frac{\cal H}{\Theta_{eq}}\ =\ \underline{0}.
\ee
From \R{I4} follows the canonical density operator of a closed 
non-isolated system in equilibrium
\bee{a48}
\varrho_{can}\ =\ \frac{1}{Z} 
\exp\Big[-\frac{{\cal H}}{k_B \Theta}_{eq}\Big],\quad
Z\ =\ \mbox{Tr}\exp\Big[-\frac{{\cal H}}{k_B \Theta}_{eq}\Big].
\ee

If we presuppose that the equilibrium environment which contacts the 
system is a heat reservoir of the thermostatic temperature $T^\Box_\odot$,
the equilibrium contact temperature $\Theta_{eq}$ of the system 
is replaced by $T^\Box_\odot$
\bee{48a}
\Theta_{eq}\ =\ T^\Box_\odot,
\ee
representing an additional equilibrium condition.

The micro-canonical and the canonical equilibrium density operators, \R{5.1bx} and
\R{a48}, are derived by a pure phenomenological argumentation: starting with the
entropy production in isolated and non-isolated systems \R{*76}, vanishing entropy
production \R{81a}$_2$ follows from the equilibrium conditions \R{+81}. Considering a process
which brings the system off equilibrium, \R{81c}, results in equilibrium distributions
\R{+82} and \R{+83}, in isolated and non-isolated closed systems.

\subsection {Constitutive equations}
\subsubsection{Heat- and entropy-exchange}

According to the constitutive equation \R{+25}, the heat exchanges \R{+59} and
\R{+60} result in
\bey\nonumber
-\st{\td}{Q}{^{12}_{ex}} &=& \st{\td}{Q}{^{1}_{ex}}+\st{\td}{Q}{^{2}_{ex}}
-\st{\td}{Q}{_{ex}}\ =
\\ \label{I4a}
&=&\kappa{^1_{ex}}\Big(\frac{1}{\Theta^1}-\frac{1}{T^\Box}\Big)
+\kappa{^2_{ex}}\Big(\frac{1}{\Theta^2}-\frac{1}{T^\Box}\Big)
-\kappa{_{ex}}\Big(\frac{1}{\Theta}-\frac{1}{T^\Box}\Big),
\\ \nonumber
-\st{\td}{Q}{^{12}_{int}} &=& \st{\td}{Q}{^{1}_{int}}+\st{\td}{Q}{^{2}_{int}}\ =\ 
\\ \label{I4b}
&=&\kappa{^1_{int}}\Big(\frac{1}{\Theta^1}-\frac{1}{\Theta^{12}}\Big)
+\kappa{^2_{int}}\Big(\frac{1}{\Theta^2}-\frac{1}{\Theta^{12}}\Big).
\eey
From \R{I4a} follows
\bee{I4c}
-\st{\td}{Q}{^{12}_{ex}}\ =\ \frac{1}{T^\Box}\Big(\kappa_{ex}
-\kappa{_{ex}^1}-\kappa{_{ex}^2}\Big)+\frac{\kappa{_{ex}^1}}{\Theta^1}
+\frac{\kappa{_{ex}^2}}{\Theta^2}-\frac{\kappa{_{ex}}}{\Theta}.
\ee
Because $\st{\td}{Q}{^{12}_{ex}}$ and the $\kappa_{ex}$
do not depend on $T^\Box$, we obtain finally by taking \R{38b4} into account
\bee{I4d}
-\st{\td}{Q}{^{12}_{ex}}\ =\ \kappa{^1_{ex}}\Big(\frac{1}{\Theta^1}-\frac{1}{\Theta}\Big)
+\kappa{^2_{ex}}\Big(\frac{1}{\Theta^2}-\frac{1}{\Theta}\Big)\ =\ 
-\mbox{Tr}\Big({\cal H}^{12}\ro_{ex}\Big).
\ee

Consequently, the heat exchanges are additive, if $\st{\td}{Q}{^{12}_{ex}}$ and
$\st{\td}{Q}{^{12}_{int}}$ which depend on the quantum mechanical interaction
vanish. Beyond that, the contact temperatures and the heat conductivities are related
to the quantum mechanical interaction by \R{I4b}, \R{38b2} and \R{I4d}$_2$,
thus demonstrating the dependence of the constitutive equations on the quantum
mechanical background.

\subsubsection{Propagators\label{CE}}

Thus far, the time rates of the weights
$\{\st{\td}{p}_{kl}\}$ of the density operator are unknown, and
they need equations for their determination. Three aspects are of interest: what quantities
determine these time rates, on what way come different materials into play and what is
the difference between the $\{\st{\td}{p}\!{_{kl}^{ex}}\}$ and
$\{\st{\td}{p}\!{_{kl}^{iso}\}}$ in non-isolated and isolated systems ?

As in classical non-equilibrium thermodynamics, the entropy production
in quantum thermodynamics has also the typical form of a product of 
``fluxes'' $\st{\td}{\mathbf{p}}$ and ``forces'' ${\mathbf{f}}$ according to \R{*76}.
The term $\st{\td}{\mathbf p}{^{iso}}\cdot{\mathbf f}{^I}$ belongs to an isolated
system, whereas the term $\st{\td}{\mathbf p}\cdot{\mathbf f}$ to a non-isolated
closed system. Constitutive equations according to the usual scheme \R{+24} cannot be used
because the equilibrium distributions ${\mathbf f}^{I}=\mvec{0}$ and  ${\mathbf f}=\mvec{0}$
do not induce the vanishing of the propagators due to their missing sufficiency for equilibrium.
Therefore we start out with \R{IV4}$_3$ and \R{58c}$_3$
\bee{I5}
\st{\td}{\mathbf p}\!{^{iso}}\cdot{\mathbf f}^{II}\ =\ 0,\qquad   
\st{\td}{\mathbf p}\!{^{ex}}\cdot{\mathbf f}^{II}\ =\ \Xi_{ex},
\ee
and we accept the
\vspace{.3cm}\newline
$\blacksquare$\underline{Setting VII:}
\bee{I6}
\st{\td}{\mathbf p}{^{iso}}\ \mbox{and}\ \st{\td}{\mathbf p}{^{ex}}\ 
\mbox{depend on}\ {\mathbf f}{^{II}},\quad
\hspace{2.7cm}\blacksquare\hspace{-2.7cm}
\ee
that means according to \R{+34}, $\st{\td}{\mathbf p}{^{iso}}$ and $\st{\td}{\mathbf p}{^{ex}}$
depend on the total Hamiltonian and on the contact
temperature of the corresponding undecomposed system. Beyond that, the choice of the
constitutive equations have to satisfy \R{+81} in equilibrium. Consequently, we introduce
matrices ${\bf A}$ and ${\bf B}$ creating the constitutive equations of the propagators
\bee{I7}
\st{\td}{\mathbf p}\!{^{iso}}\ =\ {\bf A}(\mvec{z}^{iso},{\mathbf f}{^{II}})\cdot
{\mathbf f}{^{II}},
\quad{\bf A}^\top = -{\bf A},\qquad\quad
\st{\td}{\mathbf p}\!{^{ex}} =\ {\bf B}(\mvec{z}^{ex},{\mathbf f}{^{II}})\cdot{\mathbf f}{^{II}}
\ee
according to \R{I5}$_1$. The $\mvec{z}^{iso}$ and $\mvec{z}^{ex}$ take the
equilibrium in isolated and non-isolated closed systems into account
\byy{I8}
\mvec{z}^{iso}\ :=\ \st{\td}{\mvec{a}}\!{^{12}},\Theta^1,\Theta^2,\Theta^{12},
{\mathbf f}{^{I}},&&\mvec{z}^{ex}\ :=\ \st{\td}{\mvec{a}}\!{^{1}},
\st{\td}{\mvec{a}}\!{^{2}},\Theta^1,\Theta^2,T^{\Box},{\mathbf f},\hspace{.5cm}
\\ \label{aI8}
\mvec{z}{^{iso}_{eq}}\ =\ \mvec{0},\Theta_{eq},\Theta_{eq},\Theta_{eq},\mvec{0},
&&\mvec{z}{^{ex}_{eq}}\ =\ \mvec{0},\mvec{0},T^\Box,T^\Box,T^{\Box},\mvec{0},
\\ \label{I8a}
{\bf A}_{eq}={\bf A}(\mvec{z}{^{iso}_{eq}},{\mathbf f}{^{II}})={\bf 0},&& {\bf B}_{eq}={\bf B}(\mvec{z}{^{ex}_{eq}},{\mathbf f}{^{II}})={\bf 0}.
\eey

As the equilibrium conditions \R{*77} and \R{*77a} demonstrate, the equilibrium
distributions \R{+82} and \R{+83} are only necessary, but not sufficient for
equilibrium. This fact can be represented by
\bee{I9}
\mvec{z}^{iso}_\#\ =\ \st{\td}{\mvec{a}}\!{^{12}},\Theta^1,\Theta^2,\Theta^{12},
\mvec{0},\quad\mvec{z}^{ex}_\#\ =\ \st{\td}{\mvec{a}}\!{^{1}},
\st{\td}{\mvec{a}}\!{^{2}},\Theta^1,\Theta^2,T^{\Box},\mvec{0}.
\ee
Consequently, the non-vanishing propagators in \R{81c} become by use of \R{I7}
\bee{I10}
\st{\td}{\mathbf p}\!{^{iso}_\#}\ =\ {\bf A}(\mvec{z}^{iso}_\#)\cdot
{\mathbf f}{^{II}},\qquad
\st{\td}{\mathbf p}\!{^{ex}_\#} =\ {\bf B}(\mvec{z}^{ex}_\#)
\cdot{\mathbf f}{^{II}},
\ee
in presence of the equilibrium distributions ${\mathbf f}^{I}=\mvec{0}$ and
 ${\mathbf f}=\mvec{0}$.

The ${\bf A}$ and ${\bf B}$ have to satisfy the following relations according to
\byy{I11}
\mbox{\R{*76}$_4$}\longrightarrow\quad\Sigma &=&
-{\mathbf f}^{I}\cdot{\bf A}\cdot{\mathbf f}^{II}\ \geq\ 0,
\\ \label{I12}
\mbox{\R{I5}$_2$}\longrightarrow\quad\Xi &=&
{\mathbf f}^{II}\cdot{\bf B}\cdot{\mathbf f}^{II}\ =\ \Xi_{ex},
\\ \label{I13}
\mbox{\R{+31}$_2$}\longrightarrow\quad\st{\td}{S} &=&
-{\mathbf f}^{I}\cdot({\bf A}+{\bf B})\cdot{\mathbf f}^{II},
\\ \label{I14}
\mbox{\R{IV3}$_3$}\longrightarrow\quad 0 &=& 
{\mathbf f}\cdot{\bf B}\cdot{\mathbf f}^{II}.
\eey
The relations \R{I11} and \R{I14} are constraints which have to be satisfied by the
constitutive matrices ${\bf A}$ and ${\bf B}$.

\subsection{Special processes}

\subsubsection{Adiabatic processes}

We distinguish two kinds of adiabatic processes: according to \R{+59}, we define 
\byy{P1}
\mbox{weak adiabatic:}&&\qquad \st{\td}{Q}_{ex}\ =\ 0,\quad 
\st{\td}{Q}{_{ex}^1}\ =\ -\st{\td}{Q}{_{ex}^2}\ \neq\ 0,
\\ \label{P2}
\mbox{strong adiabatic:}&&\qquad\st{\td}{Q}_{ex}\ =\ 0,\quad 
\st{\td}{Q}{_{ex}^1}\ =\ \st{\td}{Q}{_{ex}^2}\ =\ 0.
\eey
In both cases, $\st{\td}{Q}{_{ex}^{12}}$ vanishes according to \R{+59} and \R{I4d}
results in
\bee{P3}
\kappa{^1_{ex}}\Big(\frac{1}{\Theta^1}-\frac{1}{\Theta}\Big)\ =\
-\kappa{^2_{ex}}\Big(\frac{1}{\Theta^2}-\frac{1}{\Theta}\Big),
\ee
that means, the contact temperatures depend in the weak adiabatic case on
$\kappa{^1_{ex}}$ and $\kappa{^2_{ex}}$, whereas in the strong adiabatic case
$\Theta=T^\Box=\Theta^1=\Theta^2$ is valid.

\subsubsection{Reversible processes}

Reversible processes are defined by vanishing entropy production without equilibrium,
according to \R{*76}
\bee{P4} 
\Big(\Sigma_{rev}\ =\ 
-\st{\td}{\mathbf p}\!{^{iso}_{rev}}\cdot{\mathbf f}{^I_{rev}}\ =\ 0\Big)
\quad\wedge\quad
\Big(\st{\td}{\mathbf p}\!{^{iso}_{rev}}\ \neq\ \mvec{0}\quad\vee\quad
{\mathbf f}{^I_{rev}}\ \neq\ \mvec{0}\Big).
\ee
According to \R{I7}$_1$,
\bee{P5} 
{\mathbf f}{^I_{rev}}\cdot{\bf A}(\mvec{z}{^{iso}_{rev}},{\mathbf f}{^{II}})\cdot
{\mathbf f}{^{II}}\ =\ 0
\ee
represents the {\em condition of reversibility}. There are two cases for satisfying this
condition
\bee{P6}
\Big({\mathbf f}{^I_{rev}}\cdot{\bf A}(\mvec{z}{^{iso}_{rev}},{\mathbf f}{^{II}})\
=\ \mvec{0}\quad\vee\quad
{\mathbf f}{^I_{rev}}\ =\ {\mathbf f}{^{II}}\cdot{\bf C}\Big),
\quad{\bf C}\cdot{\bf A}=-{\bf A}^\top\cdot{\bf C}^\top.
\ee
The special case
\bee{P7}
{\bf C}\ =\ -{\bf 1}\ \longrightarrow\ 
{\mathbf f}{^I_{rev}}\ =\ -{\mathbf f}{^{II}}\ \longrightarrow\ {\mathbf f}{_{rev}}\
=\ \mvec{0}
\ee
belongs to the equilibrium of non-isolated closed systems \R{+83}. Be aware that the
${\mathbf f}$ are quantum mechanically defined items according to \R{+32}, \R{+34}
and \R{+36}.

\section{Decomposed Systems\label{DS}}

According to the definition given in the beginning of sect.\ref{CS}, the density ope\-rator
$\varrho_{com}$ of the compound description is replaced by the two partial density
operators --$\varrho^1\ \mbox{and}\ \varrho^2$ of the two sub-systems
\#1 and \#2 of the bipartite system according to \R{+29} and \R{+28}
\byy{59a}
\mbox{Tr}^2\varrho_{com} &=:&
\varrho^1\ =\ \sum_{kj} p_{kj}|\Psi_1^k\rangle\langle\Psi^k_1|,
\\ \label{59b}
\mbox{Tr}^1\varrho_{com} &=:&
\varrho^2\ =\ \sum_{jl} p_{jl}|\Psi_2^l\rangle\langle\Psi^l_2|,
\\ \label{59c}
\mbox{Tr}^2\ro_{com} &=:&
\ro{^1}\ =\ \sum_{kj}\st{\td}{ p}_{kj}|\Psi_1^k\rangle\langle\Psi^k_1|,
\\ \label{59d}
\mbox{Tr}^1\ro_{com} &=:&
\ro{^2}\ =\ \sum_{jl}\st{\td}{p}_{jl}|\Psi_2^l\rangle\langle\Psi^l_2|.
\eey
The decomposition of the Hamiltonian is the same as for the compound
description \R{+37}.

The traces of the density operators and of the propagators are by
taking
\R{59a} to \R{59d} into account
\byy{a61}
1\ =\ \mbox{Tr}\varrho_{com}\ =\ \mbox{Tr}^1{\varrho}{^1} \ =\ 
\mbox{Tr}^2{\varrho}{^2}\ =\ \sum_{ij}p{_{ij}}, 
\\ \label{61a}
0\ =\ \mbox{Tr}\ro_{com}\ =\ \mbox{Tr}^1{\ro}{^1} \ =\ 
\mbox{Tr}^2{\ro}{^2}\ =\ \sum_{ij} \st{\td}{p}_{ij}.
\vspace{.3cm}\eey

\subsection{Partial entropies\label{PE}}

Because in decomposed systems, the density operator $\varrho_{com}$ of the compound
system is raplaced by those of the bipartite system --\R{59a} and \R{59b}-- we are able to define partial entropies of the sub-systems.\footnote{Thermodynamical
quantities of the bipartite system are written in bold face.}
Starting with the entropy of the compound system \R{+14}
\bee{+111}
S(\varrho_{com})\ =\ -k_B\mbox{Tr}(\varrho_{com}\ln\varrho_{com})
\ee
we define
\bee{+112}
{\bf S}_{1}(\varrho^{1})\ :=\ -k_B\mbox{Tr}^1(\varrho^{1}\ln\varrho^{1}),\qquad
{\bf S}_{2}(\varrho^{2})\ :=\ -k_B\mbox{Tr}^2(\varrho^{2}\ln\varrho^{2})
\ee
by using the partial density operators \R{59a} and \R{59b}.
According to \R{3.3} in sect.\ref{TR}, these partial entropies result in
\bee{+113}
{\bf S}_{1}(\varrho^{1})\ =\ -k_B\mbox{Tr}(\varrho_{com}\ln\varrho^{1}),\qquad
{\bf S}_{2}(\varrho^{2})\ =\ -k_B\mbox{Tr}(\varrho_{com}\ln\varrho^{2}).
\ee
A comparison of \R{+113} with \R{+111} depicts that the partial entropies are not additive with respect to the entropy of the compound system
\bee{+114}
{\bf S}_1 + {\bf S}_2 - S\ =\ -k_B\mbox{Tr}\Big\{\varrho_{com}\Big(\ln(\varrho^{1}
\varrho^{2})-\ln\varrho_{com}\Big)\Big\},
\ee
because the density operator of the compound system does not decompose in general.

Using Klein's inequality \C{KLEIN}
\bee{+115}
\mbox{Tr}(A\ln B)-\mbox{Tr}(A\ln A)\ \leq\ \mbox{Tr}B - \mbox{Tr}A, 
\ee
we obtain according to \R{+115} and \R{a61}
\byy{+116}
\mbox{Tr}\Big\{\varrho_{com}\Big(\ln(\varrho^{1}\varrho^{2})
-\ln\varrho_{com}\Big)\Big\}\ \leq\ \mbox{Tr}(\varrho^{1}\varrho^{2})
-\mbox{Tr}\varrho_{com}\ =\ 0,
\\ \label{+116a}
\mbox{Tr}(\varrho^{1}\varrho^{2})\ = \
\mbox{Tr}^1\mbox{Tr}^2(\varrho^{1}\varrho^{2})\ =\ 1,\hspace{1cm}
\eey
and \R{+114} results in
\bee{+117}
{\bf S}_1 + {\bf S}_2\ \geq\ S.
\ee
If the entropy of the compound system does not decompose into the partial entropies of the
decomposed system, it is according to \R{+117} smaller than the sum of the partial entropies.
The descriptions of a system as a compound one or as a decomposed one are different, a fact
which is denoted as {\em compound deficiency} \C{MUBE04}.

\subsection{Equations of motion\label{EOM}}

Using the calculation rules of app.\ref{TR} and the decomposition of the 
Hamiltonian \R{+37}, the modified von Neumann equation \R{+43z}
\bee{+43}
\st{\td}{\varrho}_{com}\ =\ 
-\frac{i}{\hbar} \Big[({\cal H}^1+{\cal H}^2+{\cal H}^{12}),\varrho_{com}\Big]+\ro_{com}
\ee
results in two equations of motion for the traced density operators of the sub-systems
by taking \R{59s} into account
\bey\nonumber
\st{\td}{\varrho}{^1}\ :=\ \mbox{Tr}^2\st{\td}{\varrho}_{com} &=& 
-\frac{i}{\hbar}\mbox{Tr}^2 \Big[{\cal H}^1,\varrho_{com}\Big]
-\frac{i}{\hbar}\mbox{Tr}^2 \Big[{\cal H}^{12},\varrho_{com}\Big]
+\mbox{Tr}{^2\!\ro_{com}}\ =
\\ \label{+44} 
&=& -\frac{i}{\hbar}\Big[{\cal H}^1,\varrho^1\Big]
-\frac{i}{\hbar}\mbox{Tr}^2 \Big[{\cal H}^{12},\varrho_{com}\Big]+\ro{^1},
\\  \nonumber
\st{\td}{\varrho}{^2}\ :=\ \mbox{Tr}^1\st{\td}{\varrho}_{com} &=& 
-\frac{i}{\hbar}\mbox{Tr}^1 \Big[{\cal H}^2,\varrho_{com}\Big]
-\frac{i}{\hbar}\mbox{Tr}^1 \Big[{\cal H}^{12},\varrho_{com}\Big]
+\mbox{Tr}^1\ro_{com}\ =
\\ \label{+45}
&=&-\frac{i}{\hbar} \Big[{\cal H}^2,\varrho^2\Big]
-\frac{i}{\hbar}\mbox{Tr}^1 \Big[{\cal H}^{12},\varrho_{com}\Big]
+\ro{^2}.
\eey

As \R{+44} and \R{+45} depict, the traced operators $\st{\td}{\varrho}\!{^1}$
and $\st{\td}{\varrho}\!{^2}$ include as expected the interaction Hamiltonian
${\cal H}^{12}$ and beyond that, also the density operator $\varrho_{com}$
\R{+29}$_1$ of the compound system.
That means, traced operators, although acting on the sub-space of a sub-system
include quantities belonging to the bipartite compound system on the whole.
The density operators $\varrho^1$ and $\varrho^2$
and the propagators $\ro\!{^1}$ and $\ro\!{^2}$
of the sub-systems \#1 and \#2 result from un\-in\-ver\-tible tracing 
according to \R{59a} to \R{59d}:
\bee{80x}
\varrho_{com}\ \longmapsto\ (\varrho^1,\ \varrho^2)\
\longrightarrow\hspace{-.5cm}|\hspace{.5cm} \varrho_{com},
\qquad
\ro_{com}\ \longmapsto\ (\ro{^1},\ \ro{^2})\
\longrightarrow\hspace{-.5cm}|\hspace{.5cm} \ro_{com}.
\ee
That means, the knowledge of $\varrho^1$ and $\varrho^2$ does not replace
$\varrho_{com}$.

If we introduce
\bee{+45a}
\st{\trido}{\varrho}\!{^1}\ :=\ -\frac{i}{\hbar}\mbox{Tr}^2 \Big[{\cal H}^{12},\varrho_{com}\Big]+\ro{^1},\quad
\st{\trido}{\varrho}\!{^2}\ :=\ -\frac{i}{\hbar}\mbox{Tr}^1 \Big[{\cal H}^{12},\varrho_{com}\Big]+\ro{^2},
\ee
the equations of motion \R{+44} and \R{+45} result in
\bee{+45b}
\st{\td}{\varrho}{^1}\ =\  -\frac{i}{\hbar}\Big[{\cal H}^1,\varrho^1\Big]
+\st{\trido}{\varrho}\!{^1},\quad
\st{\td}{\varrho}{^2}\ =\  -\frac{i}{\hbar}\Big[{\cal H}^2,\varrho^2\Big]
+\st{\trido}{\varrho}\!{^2}.
\ee
The shape of the partial equations of motion \R{+45b} of the sub-sytems is identical
with that of the compound system \R{+43z} except for $\ro$ is replaced by
$\st{\trido}{\varrho}\!{^1}$ and $\st{\trido}{\varrho}\!{^2}$.

\subsection{The exchanges\label{EXC}}

\subsubsection{Power exchange}

The Hamiltonian and its decomposition \R{+37} is identical to those of compound
systems \R{38d2} and \R{38d3}. Consequently, also the power exchanges \R{38d4}
and \R{38d5} remain unchanged by the transfer to compound systems
\byy{L1}
&&\w{_ 1^{ex}} \equiv\ \st{\td}{W}\!{{^1}\!\!_{ex}} =
 \mbox{Tr}^1 \mbox{Tr}^2
\Big(\frac{{\partial\cal H}^1}{\partial\mvec{a}^1}\varrho_{com}\Big)
\cdot\st{\td}{\mvec{a}}\!{^1}= 
\mbox{Tr}^1\Big(\frac{{\partial\cal H}^1}{\partial\mvec{a}^1}\varrho^1\Big)
\cdot\st{\td}{\mvec{a}}\!{^1},
\\ \label{L2}
&&\w{_ 1^{int}} \equiv\ \st{\td}{W}\!{{^1}\!\!_{int}} =
\mbox{Tr}^1\mbox{Tr}^2\Big(\frac{{\partial\cal H}^1}{\partial\mvec{a}^{12}}\varrho_{com}\Big)
\cdot\st{\td}{\mvec{a}}\!{^{12}} = \mbox{Tr}^1\Big(\frac{{\partial\cal H}^1}{\partial\mvec{a}^{12}}\varrho^1\Big)
\cdot\st{\td}{\mvec{a}}\!{^{12}},\hspace{.5cm}
\\ \label{L3}\
&&\w{_ {12}^{int}} \equiv\  \st{\td}{W}\!{{^{12}}\!\!_{int}} =
\mbox{Tr}\Big(\frac{{\partial\cal H}^{12}}{\partial\mvec{a}^{12}}\varrho_{com}\Big)
\cdot\st{\td}{\mvec{a}}\!{^{12}}.
\eey
Also the relations of external and internal power exchange \R{38d6} and \R{38e1}
remain their validity for decomposed described systems. The concept of an inert partition
is independent of a compound or decomposed description: 
$\w{_ {12}^{int}} \equiv 0$.

\subsubsection{Heat- and entropy-exchange}

First of all, the relations \R{38b} and \R{+59} to \R{+62} remain valid in decomposed
systems because they are of thermodynamical origin and thus independent of the
quantum mechanical description of the system. As the power exchanges, also the heat
exchanges are independently defined of the system's description. Consequently, we have
according to \R{38a} and \R{38a1}
\byy{L3a}
\q_A &:=& 
\mbox{Tr}^A({\cal H}^A\st{\td}{\varrho}{^A})\ =\ 
\mbox{Tr}({\cal H}^A\st{\td}{\varrho}{_{com}})\ =\
\st{\td}{Q}{^A},\quad\mbox{A=1,2.}
\\ \label{L3a1}
{\bf\Xi}_A &:=& 
\mbox{Tr}^A\Big(\frac{{\cal H}^A}{\Theta^A}\st{\td}{\varrho}{^A}\Big)\ =\ 
\mbox{Tr}\Big(\frac{{\cal H}^A}{\Theta^A}\st{\td}{\varrho}{_{com}}\Big)\ =\
\Xi{^A},\quad\mbox{A=1,2.}
\eey
According to \R{38b3}, this heat  exchange has to be split into its internal and external
part. Taking \R{38b2} and \R{38b4} into account, we obtain
\byy{L3b}
&&\q{_A^{int}}\ :=\
-\frac{i}{\hbar}\mbox{Tr}^A\Big({\cal H}^A\mbox{Tr}^B
\Big[{\cal H},\varrho_{com}\Big]\Big)
+\mbox{Tr}^A\Big({\cal H}^A\ro{_{iso}^A}\Big),\
\\ \label{L3c}
&&\q{_A^{ex}}\ :=\ \mbox{Tr}^A\Big({\cal H}^A\ro{_{ex}^A}\Big),
 \qquad\qquad\qquad
\mbox{A,B=1,2,\ A$\neq$B},
\\ \label{L3d}
&&\q{_{12}^{int}}\ =\ \st{\td}{Q}{^{12}_{int}},
\quad\q{_{12}^{ex}}\ =\ \st{\td}{Q}{^{12}_{ex}}
\eey

As the heat exchanges, also the entropy exchanges are identical in compound and
decomposed description
\byy{L3e}
&&{\bf\Xi}{_A^{int}}\ :=\
-\frac{i}{\hbar}\mbox{Tr}^A\Big(\frac{{\cal H}^A}{\Theta^A}\mbox{Tr}^B
\Big[{\cal H},\varrho_{com}\Big]\Big)
+\mbox{Tr}^A\Big(\frac{{\cal H}^A}{\Theta^A}\ro{_{iso}^A}\Big),\
\\ \label{L3f}
&&{\bf\Xi}{_A^{ex}}\ :=\
\mbox{Tr}^A\Big(\frac{{\cal H}^A}{\Theta^A}\ro{_{ex}^A}\Big),
\qquad\qquad\qquad
\mbox{A,B=1,2,\ A$\neq$B},
\\ \label{L3g}
&&{\bf\Xi}{_{12}^{int}}\ =\ \Xi{^{12}_{int}},
\quad{\bf\Xi}{_{12}^{ex}}\ =\ \Xi{^{12}_{ex}}.
\eey

\subsection{Entropy rate and contact temperature\label{ERCT}}

Starting with \R{+113}
\bee{L4}
{\bf S}_A\ =\ -k_B\mbox{Tr}(\varrho_{com}\ln\varrho^A)\ =\
-k_B\mbox{Tr}^A(\varrho^A\ln\varrho^A)
\ee
results by taking \R{+44} into account
\bey\nonumber
\st{\td}{{\bf S}}_A &=& -k_B\mbox{Tr}^A(\st{\td}{\varrho}{^A}\ln\varrho^A)\ =\
\\ \nonumber
 &=&
-k_B\mbox{Tr}^A\Big\{\Big(-\frac{i}{\hbar}\Big[{\cal H}^A,\varrho^A\Big]
-\frac{i}{\hbar}\mbox{Tr}^B\Big[{\cal H}^{12},\varrho_{com}\Big]+\ro{^A}\Big)
\ln\varrho^A\Big\}=\hspace{1cm}
\\ \nonumber
&=&-k_B\mbox{Tr}^A\Big\{\Big(-\frac{i}{\hbar}\mbox{Tr}^B\Big[{\cal H}^{12},\varrho_{com}\Big]+\ro{^A}\Big)\ln\varrho^A\Big\}\ =\
\\ \nonumber
&=&k_B\mbox{Tr}\Big(\frac{i}{\hbar}\Big[{\cal H}^{12},\varrho_{com}\Big]
\ln\varrho^A\Big)-k_B\mbox{Tr}^A(\ro{^A}\ln\varrho^A)\ =\
\\ \label{L5}
&=& -k_B\mbox{Tr}\Big\{\Big(\ro_{com}-\frac{i}{\hbar}\Big[{\cal H}^{12},\varrho_{com}\Big]\Big)\ln\varrho^A\Big\}.
\eey
Summing up the entropy time rates of the sub-systems,
we obtain from \R{L5} and \R{+41}
\bey\nonumber
&&\s_1+\s_2 - \st{\td}{S}\ =\
\\ \nonumber
&&=\ -k_B\mbox{Tr}\Big\{\Big(\ro_{com}-\frac{i}{\hbar}\Big[{\cal H}^{12},\varrho_{com}\Big]\Big)\ln(\varrho^1\varrho^2)-\ro_{com}\ln\varrho_{com}\Big\}\ =\ 
\\ \label{+131}
&&=\ -k_B\mbox{Tr}\Big\{-\frac{i}{\hbar}\Big[{\cal H}^{12},\varrho_{com}\Big]\ln(\varrho^1\varrho^2)+\ro_{com}\Big(\ln(\varrho^1\varrho^2)-\ln\varrho_{com}\Big)\Big\},\hspace{1cm}
\eey
that means, the sum of the entropy rates of the sub-sytems is different from the entropy
rate of the undecomposed system. The same non-additivity which appears for the entropy
itself according to \R{+117} is called compound deficiency. Some more details are
discussed in sect.\ref{CD}.   

According to \R{+20}$_1$ and taking \R{L3a} and \R{L3a1} into account,
the entropy production of the sub-system $\#A$ is defined by
\bee{L5a}
{\bf\Sigma}_ A\ :=\ \s_A - {\bf\Xi}_A\ =\ \s_A - \frac{\q_ A}{\Theta_A}\ =\ 
\s_A- \frac{\st{\td}{Q}{^A}}{\Theta_A}. 
\ee
Consequently, the quantum mechanical expression of the contact temperature $\Theta_A$
of sub-system $\#A$ results in
\bee{L6}
\frac{1}{\Theta^A}\ =\ \frac{\s_A}{\q_ A}\ =\
\frac{ -k_B\mbox{Tr}\Big\{\Big(\ro_{com}-\frac{i}{\hbar}\Big[{\cal H}^{12},\varrho_{com}\Big]\Big)\ln\varrho^A\Big\}}{-\frac{i}{\hbar}\mbox{Tr}\Big({\cal H}^A\Big[{\cal H}^{12},\varrho_{com}\Big]\Big)+
\mbox{Tr}\Big({\cal H}^A\ro_{com}\Big)},
\ee
by taking \R{L5}, \R{L3a}$_3$ and \R{38a} into account.

\subsection{Entropy production\label{EP}}

Inserting \R{L5}, \R{38a2} and \R{38a} into \R{L5a}, we obtain   
\bey\nonumber
{\bf\Sigma}_ A\ =\ 
-k_B\mbox{Tr}\Big\{\Big(\ro_{com}-\frac{i}{\hbar}\Big[{\cal H}^{12},\varrho_{com}\Big]\Big)\ln\varrho^A\Big\}+\hspace{3cm}
\\ \nonumber
+\frac{i}{\hbar}\mbox{Tr}\Big(\frac{{\cal H}^A}{\Theta^A}\Big[{\cal H}^{12},\varrho_{com}\Big]\Big)-
\mbox{Tr}\Big(\frac{{\cal H}^A}{\Theta^A}\ro_{com}\Big)\ =\hspace{.5cm}
\\ \nonumber
=\ -k_B\mbox{Tr}\Big\{\Big(\ln\varrho^A+\frac{{\cal H}^A}{\Theta^A}\Big)\ro_{com}
\Big\}+\hspace{4.5cm}
\\ \label{L6a}
+k_B\mbox{Tr}\Big\{\frac{i}{\hbar}\Big[{\cal H}^{12},\varrho_{com}\Big]
\Big(\ln\varrho^A+\frac{{\cal H}^A}{\Theta^A}\Big)\Big\}.
\eey
From \R{L6a} follows that vanishing entropy production is not sufficient for equilibrium.
If the density operator $\varrho^A$ of the sub-system \#A is canonical, the entropy
production vanishes in the sub-system, but $\st{\td}{\varrho}\!{^A}$ may be different from zero and no equilibrium is present. If vice-versa $\st{\td}{\varrho}\!{^A}$ vanishes,
the entropy production in the sub-system is 
according to \R{L5a}$_2$, \R{L5}$_1$ and \R{L3a}$_1$
zero, but $\varrho^A$ can be arbitrary and 
equilibrium is absent. The equilibrium conditions \R{*77} and \R{*77a} are more
farreaching than the vanishing of the entropy production.

\subsection{Compound deficiency\label{CD}}

As already discussed, heat- and entropy exchanges \R{+59} to \R{+62}, the
entropies of sub-systems \R{+117} and the entropy rates \R{+131} are not
additive, that means, summing up quantities of the 
sub-systems does not result in the corresponding quantity of the undecomposed 
system. As already mentioned, there are two reasons for this non-additivity
which is called compound deficiency: the classical decomposition of the system
and the quantum interaction between the sub-systems. This item 
will be discussed in more detail.

For elucidating the compound deficiency, we compare quantities of the compound
system with the corresponding ones of the undecomposed system.
The general definition of the compound deficiency $\boxtimes_{cd}$ of the quantity
$\boxtimes$ is \C{MUBE04}
\bee{68n}
\boxtimes_{cd}\ :=\ \boxtimes - \boxtimes^1 - \boxtimes^2\ =:\
\boxtimes - \boxtimes^{da}.
\ee
Here, $\boxtimes$ belongs to the undecomposed system, whereas $\boxtimes^1$
and $\boxtimes^2$ belong to the sub-systems of the corresponding
bipartite system, and $\boxtimes^{da}$ is
the quantity generated by decomposed additivity. In general, the
compound deficiency does not vanish. Consequently, there are two
possibilities to describe a system: as a decomposed one by 
$\boxtimes^{da}$ or as an undecomposed one by $\boxtimes$.
These descriptions are of different
information about the system, if the compound deficiency is not zero.
The answer to the question ``What is the correct entropy production of the
system?'' depends on its description chosen as decomposed or undecomposed.

According to \R{68n}, we obtain from \R{+37} the compound deficiency of
the Hamiltonian
\bee{68o}
{\cal H}_{cd}\ :=\ {\cal H} - {\cal H}^1 - {\cal H}^2\ =\ {\cal H}^{12},
\ee
which is caused by the interaction part, a fact which is obvious: the sum of the
Hamiltonians of the sub-sytems differ from the Hamiltonian of the corresponding
undecomposed system.

\subsection{Removing semi-classical description\label{RSCD}}

Up to now, the systems, described in sect.\ref{US} and sect\ref{CS}, 
are treated semi-classically, that means, the external exchange quantities are not
connected to an interaction Hamiltonian. Thus, the external heat exchange
\R{38b4} and the external
entropy exchange \R{38b5} vanish in isolated systems not by the vanishing of an   
interaction Hamiltonian, but by that of the exchange propagator \R{+52}$_1$.
This procedure is called {\em semi-classical}, because the contact temperature
\R{+23} is used as a classical quantity whose quantum-mechanical definition
\R{+77} was not taken into account (because it is unknown on this level). 
We now remove this semi-classical
description by introducing an interaction Hamiltonian describing the partition
between an undecomposed sub-system and its environment which is later on 
chosen as an equilibrium heat reservoir.

We now consider an externally isolated bipartite system which is according to \R{+50}
described by
\bee{L7}
\ro_{ex}\ \equiv\ 0\ \longrightarrow\ \ro_{com}\ \equiv\ \ro_{iso}. 
\ee
The sub-system \#2 is now the environment of sub-system \#1. The quantum
mechanical interaction between them is represented by the Hamiltonian ${\cal H}^{ia}$
For more convenience, we change the denotation of the partial Hamiltonians
\bey\nonumber
\mbox{the system:}&& {\cal H}^{1}\ \Rightarrow\ {\cal H},
\\ \nonumber
\mbox{the environment:}&& {\cal H}^{2}\ \Rightarrow\ {\cal H}^\Box,
\\ \nonumber
\mbox{the interaction:}&& {\cal H}^{12}\ \Rightarrow\ {\cal H}^{ia},
\\ \nonumber
\mbox{the total system:}&& {\cal H}\ \Rightarrow\ {\cal H}_{tot},
\\ \label{L8}
&&{\cal H}_{tot}\ =\ {\cal H} + {\cal H}^\Box + {\cal H}^{ia},
\eey
 the density operators
\bee{L9}
\mbox{Tr}^2\varrho_{com}\ =\ \varrho^1\ \Rightarrow\ \varrho\qquad
\mbox{Tr}^1\varrho_{com}\ =\ \varrho^2\ \Rightarrow\ \varrho^\Box,
\ee
 the work variables
\bee{L10}
\st{\td}{\mvec{a}}\!{^{1}}\ \Rightarrow\ \mvec{0},\qquad
\st{\td}{\mvec{a}}\!{^{2}}\ \Rightarrow\ \mvec{0},\qquad
\st{\td}{\mvec{a}}\!{^{12}}\ \Rightarrow\ \st{\td}{\mvec{a}}\!{^{ia}},
\ee
and the contact temperatures
\bee{L11}
\Theta^1\ \Rightarrow\ \Theta,\qquad\Theta^2\ \Rightarrow\ \Theta^\Box,\qquad
\Theta^{12}\ \Rightarrow\ \Theta^{ia}.
\ee

System and its environment are each undecomposed sub-systems of the bipartite system
whose environment is not defined and excluded by $\ro_{ex}\equiv 0$. In more detail,
we obtain after having inserted the changed denotations:

\subsubsection{Equations of motion}
\vspace{-.5cm}
\bey\nonumber
&&\mbox{from \R{+43}:}
\\ \label{L12}
&&\st{\td}{\varrho}_{com}\ =\ 
-\frac{i}{\hbar} \Big[({\cal H}+{\cal H}^\Box+{\cal H}^{ia}),\varrho_{com}\Big]+\ro_{com},\quad\mbox{Tr}^{com}\ =\ \mbox{Tr}\mbox{Tr}^\Box,\hspace{.5cm}
\\ \nonumber
&&\mbox{from \R{+44} and \R{+45}:}
\\ \label{L13}
&&\st{\td}{\varrho}\ =\ -\frac{i}{\hbar}\Big[{\cal H},\varrho\Big]
-\frac{i}{\hbar}\mbox{Tr}^\Box \Big[{\cal H}^{ia},\varrho_{com}\Big]+\ro,
\\ \label{L14}
&&\st{\td}{\varrho}{^\Box}\ =\ 
-\frac{i}{\hbar} \Big[{\cal H}{^\Box},\varrho{^\Box}\Big]
-\frac{i}{\hbar}\mbox{Tr}\Big[{\cal H}^{ia},\varrho_{com}\Big]+\ro{^\Box}.
\eey

\subsubsection{Power exchange}
\vspace{-.5cm}
\bey\nonumber
&&\mbox{from \R{L2}, \R{L3} and \R{38e}}
\\ \nonumber
&&\Big\{
\mbox{Tr}\Big(\frac{{\partial\cal H}}{\partial\mvec{a}^{ia}}\varrho\Big)
+\mbox{Tr}{^\Box}\Big(\frac{{\partial\cal H}^\Box}{\partial\mvec{a}^{ia}}
\varrho{^\Box}\Big)
+\mbox{Tr}^{com}\Big(\frac{{\partial\cal H}^{ia}}{\partial\mvec{a}^{ia}}
\varrho_{com}\Big)\Big\}
\cdot\st{\td}{\mvec{a}}\!{^{ia}}=\hspace{.5cm}
\\ \label{L15}
&&=\ \w+\w\!{_ {\Box}}+\w\!{_ {ia}}\ =\ \mvec{0}.
\eey

\subsubsection{Heat exchange}
\vspace{-.5cm}
\bey \nonumber
&&\mbox{from \R{L3b}}
\\ \label{L16}
&&\q\ =\
-\frac{i}{\hbar}\mbox{Tr}\Big({\cal H}\mbox{Tr}{^\Box}
\Big[{\cal H}_{tot},\varrho_{com}\Big]\Big)
+\mbox{Tr}\Big({\cal H}\ro{^A}\Big),
\\ \label{L17}
&&\q{_\Box}\ :=\
-\frac{i}{\hbar}\mbox{Tr}{^\Box}\Big({\cal H}{^\Box}\mbox{Tr}
\Big[{\cal H}_{tot},\varrho_{com}\Big]\Big)
+\mbox{Tr}{^\Box}\Big({\cal H}{^\Box}\ro{^\Box}\Big),
\\ \label{L18}
&&\q{_{ia}}\ :=\
-\frac{i}{\hbar}\mbox{Tr}{^{com}}\Big({\cal H}^{ia}
\Big[{\cal H}_{tot},\varrho_{com}\Big]\Big)
+\mbox{Tr}{^{com}}\Big({\cal H}{^{ia}}\ro{_{com}}\Big),
\\ \nonumber
&&\mbox{from \R{+60}}
\\ \label{L19}
&&\q+\q_{\Box}+\q_{ia}\ =\ \mvec{0}.
\eey

\subsubsection{Entropy rate and production}
\vspace{-.5cm}
\bey\nonumber
&&\mbox{from \R{+131}}
\\ \nonumber
&&\s+\s_{\Box} - \st{\td}{S}\ =\
\\ \label{L20}
&&= -k_B\mbox{Tr}^{com}\Big\{\Big(\ro_{com}-\frac{i}{\hbar}\Big[{\cal H}^{ia},\varrho_{com}\Big]\Big)\ln(\varrho\varrho^\Box)-\ro_{com}\ln\varrho_{com}\Big\}. 
\\ 
&&
= -k_B\mbox{Tr}^{com}\Big\{-\frac{i}{\hbar}\Big[{\cal H}^{ia},\varrho_{com}\Big]\ln(\varrho\varrho^\square)+
\ro_{com}\Big(\ln(\varrho\varrho^\square)-\ln\varrho_{com}\Big)\Big\}.\hspace{1cm} 
\eey
The entropy rates are additive, if the quantum mechanical interaction vanishes and the density operator of the decomposed system decomposes into the product of the density
operators of the sub-systems.
\bey\nonumber
&&\mbox{from \R{L6a}}
\\ \label{L21}
&&{\bf\Sigma}_ A\ =\ 
-k_B\mbox{Tr}\Big\{\Big(\ln\varrho^A+\frac{{\cal H}^A}{\Theta^A}\Big)
\st{\td}{\varrho}_{com}\Big\}
\eey

\subsubsection{Equilibrium and heat reservoir}

A part of the necessary equilibrium conditions \R{*77} and \R{*77a} is
\bee{L22}
\st{\td}{\varrho}_{eq}\ =\ 0,\quad\st{\td}{\varrho}{_{eq}^\Box}\ =\ 0,\quad
\ro_{eq}\ =\ 0,\quad\ro{_{eq}^\Box}\ =\ 0.
\ee
According to \R{+44} and \R{+45},  the equilibrium conditions \R{L22} result in
\byy{L23}
\Big[{\cal H},\varrho_{eq}\Big]\ =\ 
-\mbox{Tr}^\Box\Big[{\cal H}^{ia},\varrho{^{eq}_{com}}\Big],
\\ \label{L24}
\Big[{\cal H}^\Box,\varrho{_{eq}^\Box}\Big]\ =\ 
-\mbox{Tr}\Big[{\cal H}^{ia},\varrho{^{eq}_{com}}\Big].
\eey

Obvious is, that $\varrho_{eq}$ and $\varrho{_{eq}^\Box}$ are not of canonical
or micro-canonical form, \R{a48} or \R{5.1bx}, because the RHSs of
\R{L23} and \R{L24} are not zero in general, except if the quantum mechanical
interaction vanishes. If there is a mutual
equilibrium of two interacting sub-systems, the density operators of these sub-systems
do not commute with the corresponding Hamiltonians, if the right-hand traces do not 
vanish. 

We now presuppose that the environment $\#^\Box$ of the system is a heat reservoir.
In this case, the density operator $\varrho{_{eq}^\Box}=\varrho_{can}$ is canonical
\R{a48} and commutes with ${\cal H}^\Box$. Consequently,
\bee{L25}
0\ =\ \mbox{Tr}\Big[{\cal H}^{ia},\varrho{^{eq}_{com}}\Big]
\ee
is valid.
Reservoir property means that the heat reservoir, during the contact with the
undecomposed system in consideration, is always in equilibrium having the 
thermostatic temperature $T^\Box$. Because it is a closed non-isolated system 
in equilibrium, the density operator of the heat reservoir is canonical for all times
\byy{L26}
\varrho^\Box_{_{eq}}(t)\  =\ \varrho_{can} = \frac{1}{Z}\exp\Big\{\!
-\frac{{\cal H}^\Box}{k_BT{^\Box_{eq}}}\Big\}\ =\ \mbox{Tr}\varrho{^{eq}_{com}}.
\eey
The equilibrium density operator $\varrho{^{eq}_{com}}$ of the undecomposed
system which belongs to the considered bipartite system has to satisfy \R{L25} and
according to \R{L23}
\bee{L27}
\Big[{\cal H},\mbox{Tr}^\Box\varrho{^{eq}_{com}}\Big]\ =\ 
-\mbox{Tr}^\Box\Big[{\cal H}^{ia},\varrho{^{eq}_{com}}\Big].
\ee
Beyond that
\bee{L28}
0\ =\ \mbox{Tr}^{com}\Big[{\cal H}^{ia},\varrho{^{eq}_{com}}\Big]
\ee
is valid according to \R{L23} and \R{L25}.

\section{Summary}

The density operator satisfying the von Neumann equation is in conventional quantum
mechanics time independently composed of pure quantum states. If this time
independence is waived, the von Neumann equation is modified by a so-called
propagator without changing the quantum mechanical dynamics. The composition
of the density operator becomes time dependent in contrast to the original von
Neumann equation. This introduction of the propagator makes possible to define
thermodynamical quantities beyond quantum mechanics.

The following non-equilibrium quantities of an undecomposed
system\footnote{undecomposed means: neither the
Hamiltonian nor the density operator of the bipartite system are decomposed according
to the sub-systems} in contact with an equilibrium environment can be defined by
combined thermodynamical and quantum mechanical settings: power- and
heat-exchange, the entropy time rate based on the Shannon entropy, entropy exchange
and production. The interaction between system and environment is preliminary
semi-classically\footnote{semi-classical means: heat- and entropy-exchange between
system and environment are classically described without a special interaction
Hamiltonian} described. Heat- and entropy-exchange depend on the propagator and on
the Hamiltonian, whereas the power exhange is independent of the propagator.
The entropy production is an internal quantity of the system, independent of the
environment, of the Hamiltonian and of the contact temperature, it depends only on the
propagator.

Taking the decomposition of the Hamiltonian into its parts according to the sub-systems
and their interaction into account, a compound system can be defined. The above
mentioned exchanges are decomposed into internal and external ones by decomposing
the propagator. As a priliminary auxiliary procedure, the interaction between system and
environment is described semi-classically. Some special cases are investigated:
vanishing quantum mechanical interaction between the sub-systems, external heat- and entropy-exchanges, equilibria and the equilibrium distributions of the isolated and
non-isolated system are derived by thermodynamical considerations. The Second Law is
taken into consideration, and a quantum theoretical expression of the contact
temperature is derived. Preliminary constitutive equations of the propagator are sketched.

The decomposition of the density operator by tracing out one or the other sub-system
makes the semi-classical description dispensable. Partial entropies and entropy rates
of the sub-systems, and equations of motion of the sub-system density operators are
introduced.
Equilibria of bipartite systems are considered, especially those for which one sub-system is a heat reservoir. Restrictions for the density operator of the corresponding
compound system in equilibrium are derived.

\section{Appendices}
\subsection{Appendix 1: Tracing\label{TR}}

Starting with a tensor of the decomposed system
\bee{3.1}
A\ =\
\sum_{kl}\sum_{pq}|\Psi_1^k\rangle|\Psi_2^l\rangle A^{pq}_{kl}\langle\Psi_2^q|
\langle\Psi_1^p|,
\ee
and its traces
\byy{3.2}
\mbox{Tr}^1 A\ =\ \sum_{lq} |\Psi_2^l\rangle \sum_m
A^{mq}_{ml}\langle\Psi_2^q|\ =:\ A^2,
\\ \label{3.2a}
\mbox{Tr}^2 A\ =\ \sum_{kp} |\Psi_1^k\rangle \sum_m
A^{pm}_{km}\langle\Psi_1^p|\ =:\ A^1.
\eey
Because partial traces commute, we obtain
\bee{59y}
\mbox{Tr}^2 A^2\ =\ \mbox{Tr}^2\mbox{Tr}^1 A \ =\ \mbox{Tr}A\ =\ 
\mbox{Tr}^1\mbox{Tr}^2 A\ =\ \mbox{Tr}^1 A^1\ =\ \sum_{mr}A^{rm}_{rm}.
\ee
\bee{3.3}
\mbox{Tr}(A^1 B)\ =\ \mbox{Tr}^1\mbox{Tr}^2(A^1B)\ =\
\mbox{Tr}^1(A^1B^1).
\ee
A detailled calculation
\bey\nonumber
\mbox{Tr}^1(A^1B)\ =\
\sum_j\sum_{sr}\sum_{kl}\sum_{pq}\hspace{6cm}
\\ \nonumber
\langle\Psi{^j_1}|\Psi{^s_1}\rangle A{^r_s}
\langle\Psi{^r_1}|\Psi{^k_1\rangle}|\Psi{^l_2\rangle}B{^{pq}_{kl}}\langle\Psi{^q_2}|
\langle\Psi{^p_1}|\Psi{^j_1}\rangle\ =\ 
\\ \label{59z}
=\ \sum_j\sum_k\sum_{ql}   A{_j^k} B{^{jq}_{kl}}|\Psi{^l_2\rangle}\langle\Psi{^q_2}|,
\eey
\bey\nonumber
\mbox{Tr}^1(BA^1)\ =\
\sum_j\sum_{kl}\sum_{pq}\sum_{sr}\hspace{6cm}
\\ \nonumber
\langle\Psi{^j_1}|\Psi{^k_1\rangle}|\Psi{^l_2\rangle}B{^{pq}_{kl}}\langle\Psi{^q_2}|\langle\Psi{^p_1}|
\Psi{^s_1}\rangle A{^r_s}\langle\Psi{^r_1}|\Psi{^j_1}\rangle\ =\ 
\\ \label{59r}
=\ \sum_j\sum_p\sum_{ql}B{^{pq}_{jl}}A{_p^j}|\Psi{^l_2\rangle}\langle\Psi{^q_2}|\ =\ 
\sum_j\sum_k\sum_{ql}B{^{jq}_{kl}}A{_j^k}|\Psi{^l_2\rangle}\langle\Psi{^q_2}|
\eey
results in
\bee{59s}
\mbox{Tr}^1[A^1,B]\ =\ 0\ \longrightarrow\ \mbox{Tr}^2\mbox{Tr}^1[A^1,B]\ =\ 
\mbox{Tr}^1[A^1,B^1]\ =\ 0.
\ee

\subsection{Appendix 2: Entropy exchange inequality\label{EEI}}

Consider an undecomposed system of the contact temperature $\Theta$ which exchanges heat
$\st{\td}{Q}\!(T^\Box)$ with its  equilibrium environment of the thermostatic temperature
$T^\Box$. According to \R{+21}, we have after the decomposition of the heat exchange into
its positive and negative parts
\byy{a1}
\st{\td}{Q}\!{^+}(T^\Box)\geq 0,\ \st{\td}{Q}\!{^-}(T^\Box)\leq 0,\qquad
\st{\td}{Q}\!{^+}(T^\Box)+\st{\td}{Q}\!{^-}(T^\Box)\ =\ \st{\td}{Q}(T^\Box),
\\ \label{a2}
\Big(\st{\td}{Q}\!{^+}(T^\Box)+\st{\td}{Q}\!{^-}(T^\Box)\Big)
\Big(\frac{1}{\Theta}-\frac{1}{T^\Box}\Big)\ \geq\ 0.\hspace{2cm}
\\ \label{a3}
\st{\td}{Q}\!{^+}(T^\Box)\Big(\frac{1}{\Theta^+}-\frac{1}{T^\Box}\Big)\ \geq\ 0,\qquad
\st{\td}{Q}\!{^-}(T^\Box)\Big(\frac{1}{\Theta^-}-\frac{1}{T^\Box}\Big)\ \geq\ 0.
\eey
The contact temperatures are independent of the thermostatic temperatures of the environment.
Consequently , we obtain from \R{a3} and  \R{a2} for the special choice
\byy{a4}
T^\Box\ \st{*}{=}\ \Theta^-\quad \longrightarrow\quad \st{\td}{Q}\!{^-}(\Theta^-)\ =\ 0:\hspace{5cm}
\\ \label{a5}
\st{\td}{Q}\!{^+}(\Theta^-)\Big(\frac{1}{\Theta^+}-\frac{1}{\Theta^-}\Big)\ \geq\ 0,\qquad
\st{\td}{Q}\!{^+}(\Theta^-)\Big(\frac{1}{\Theta}-\frac{1}{\Theta^-}\Big)\ \geq\ 0,
\eey
and
\byy{a6}
T^\Box\ \st{*}{=}\ \Theta^+\quad \longrightarrow\quad \st{\td}{Q}\!{^+}(\Theta^+)\ =\ 0:\hspace{5cm}
\\ \label{a7}
\st{\td}{Q}\!{^-}(\Theta^+)\Big(\frac{1}{\Theta^-}-\frac{1}{\Theta^+}\Big)\ \geq\ 0,\qquad
\st{\td}{Q}\!{^-}(\Theta^+)\Big(\frac{1}{\Theta}-\frac{1}{\Theta^+}\Big)\ \geq\ 0,
\eey
resulting in
\bee{a8}
\Theta^+\ \leq\ \Theta\ \leq\ \Theta^-.
\ee

Taking \R{a8} into account, we obtain
\byy{a9}
\frac{\st{\td}{Q}\!{^+}(T^\Box)}{\Theta^+}\ \geq\ 
\frac{\st{\td}{Q}\!{^+}(T^\Box)}{\Theta},\qquad
\frac{\st{\td}{Q}\!{^-}(T^\Box)}{\Theta^-}\ \geq\ 
\frac{\st{\td}{Q}\!{^-}(T^\Box)}{\Theta},
\\ \label{a10}\longrightarrow\quad
\frac{\st{\td}{Q}\!{^+}(T^\Box)}{\Theta^+}+\frac{\st{\td}{Q}\!{^-}(T^\Box)}{\Theta^-}\ 
\geq\ \frac{\st{\td}{Q}(T^\Box)}{\Theta}\ \geq\  \frac{\st{\td}{Q}(T^\Box)}{T^\Box}.
\eey

Consider a set of heat exchanges $\st{\td}{Q}\!{^j}(T^\Box)$ between a set of
sub-systems
of the contact temperatures $\Theta^j$ and an equilibrium environment of the thermostatic temperature $T^\Box$. The proof of the inequality \R{38b}$_2$ runs as follows:
\bee{a11}
\sum_j\frac{\st{\td}{Q}\!{^j}(T^\Box)}{\Theta^j}=
\sum_k\frac{\st{\td}{Q}\!{^{k+}}(T^\Box)}{\Theta^k}
+\sum_m\frac{\st{\td}{Q}\!{^{m-}}(T^\Box)}{\Theta^m}=
\frac{\st{\td}{Q}\!{^{+}}(T^\Box)}{\Theta^+}+\frac{\st{\td}{Q}\!{^{-}}(T^\Box)}{\Theta^-}.
\ee
The sum \R{a11}$_1$ is decomposed into its positive and negative parts which are
transformed by the mean value theorem
\bey\nonumber
\st{\td}{Q}\!^{+}(T^\Box)\ :=\ \sum_k\st{\td}{Q}\!^{k+}(T^\Box),\quad
\st{\td}{Q}\!^{-}(T^\Box)\ :=\ \sum_k\st{\td}{Q}\!^{k-}(T^\Box),
\\ \label{a11a}
\longrightarrow\ \Theta^+\ \mbox{and}\ \Theta^-.
\eey
 Taking \R{a10} into account, we obtain 
\bee{a12}
\sum_j\frac{\st{\td}{Q}\!{^j}(T^\Box)}{\Theta^j}\ 
\geq\ \frac{\st{\td}{Q}(T^\Box)}{\Theta}\ \geq\  \frac{\st{\td}{Q}(T^\Box)}{T^\Box}.
\ee
The inequality \R{38b}$_2$ is a special case of \R{a12}.

\end{document}